\documentclass[sigconf]{acmart}

\usepackage{xcolor}

\def\BibTeX{{\rm B\kern-.05em{\sc i\kern-.025em b}\kern-.08em
    T\kern-.1667em\lower.7ex\hbox{E}\kern-.125emX}}

\usepackage{hyperref}
\hypersetup{hidelinks}
\usepackage{algorithmic}
\usepackage{algorithm}
\usepackage{amsmath,amsfonts}
\usepackage{mathtools}
\usepackage{stmaryrd}
\usepackage{algorithmic}
\usepackage{graphicx}
\usepackage{textcomp}
\usepackage{xcolor}
\usepackage{bm}
\usepackage{adjustbox}
\usepackage{tabularx}
\usepackage{float}
\usepackage{colortbl}
\usepackage{booktabs}
\usepackage{multirow}
\usepackage{graphicx}
\usepackage{xspace}
\usepackage[most]{tcolorbox}
\usepackage{balance}
\usepackage[normalem]{ulem}
\usepackage{tabularray}
\usepackage{pgfplots}
\usepackage{threeparttable}
\usepackage{tikz}
\usepackage{enumitem}
\usepackage{pifont}
\usepackage{pgfplots}
\usetikzlibrary{pgfplots.groupplots}
\usepackage{color}
\usepackage{colortbl}

\pgfplotsset{compat=1.18}
\useunder{\uline}{\ul}{}

\newcommand{\greyboxb}[2]{
\vspace{0.05cm}
    \begin{tcolorbox}[
        left=2pt, right=2pt, top=2pt, bottom=2pt,
        boxrule=0.2mm,
        leftrule=2mm,
        arc=0mm,
        colframe=black!40!white,
        colback=black!5!white,
        colbacktitle=black!50!white
    ]
    \textbf{#1}{#2}
    \end{tcolorbox}
\vspace{0.05cm}
}

\renewcommand\footnotetextcopyrightpermission[1]{}
\setcopyright{none}

\newcommand{\phead}[1]{\vspace{1mm} \noindent {\bf #1}}

\newcommand{\toolname}{\textsc{LiveFuzz}}

\definecolor{darkgreen}{rgb}{0.0, 0.5, 0.0}

\newboolean{showcomments}
\setboolean{showcomments}{true}
\ifthenelse{\boolean{showcomments}}
 { \newcommand{\mynote}[2]{
      \fbox{\bfseries\sffamily\scriptsize#1}
        {\small$\blacktriangleright$\textsf{\emph{#2}}$\blacktriangleleft$}}}
        { \newcommand{\mynote}[2]{}}

\begin{document}
\title{Triggering and Detecting Exploitable Library Vulnerability from the Client by Directed Greybox Fuzzing} 

\author{Yukai Zhao}
\affiliation{    
  \institution{Zhejiang University}
  \city{Hangzhou}
  \country{China}
}
\email{yukaizhao2000@zju.edu.cn}

\author{Menghan Wu}
\affiliation{%
  \institution{Zhejiang University}
  \city{Hangzhou}
  \country{China}
}
\email{menghanwu@zju.edu.cn}

\author{Xing Hu}
\authornote{Corresponding Author}
\affiliation{
  \institution{Zhejiang University}
  \city{Hangzhou}
  \country{China}
}
\email{xinghu@zju.edu.cn}

\author{Shaohua Wang}
\affiliation{    
  \institution{Central University of Finance and Economics}
  \city{Beijing}
  \country{China}
}
\email{davidshwang@ieee.org}

\author{Meng Luo}
\affiliation{    
  \institution{Zhejiang University}
  \city{Hangzhou}
  \country{China}
}
\email{meng.luo@zju.edu.cn}

\author{Xin Xia}
\affiliation{    
  \institution{Zhejiang University}
  \city{Hangzhou}
  \country{China}
}
\email{xin.xia@acm.org}


\begin{abstract}

Developers utilize third-party libraries to improve productivity, which also introduces potential security risks.
Existing approaches generate tests for public functions to trigger library vulnerabilities from client programs, yet they depend on proof-of-concepts (PoCs), which are often unavailable.
In this paper, we propose a new approach, {\toolname}, based on directed greybox fuzzing (DGF) to detect the exploitability of library vulnerabilities from client programs without PoCs.
{\toolname} exploits a target tuple to extend existing DGF techniques to cross-program scenarios.
Based on the target tuple,  {\toolname} introduces a novel Abstract Path Mapping mechanism to project execution paths, mitigating the preference for shorter paths.
{\toolname} also proposes a risk-based adaptive mutation to mitigate excessive mutation.
To evaluate {\toolname}, we construct a new dataset including 61 cases of library vulnerabilities exploited from client programs.
Results show that {\toolname} increases the number of target-reachable paths compared with all baselines and improves the average speed of vulnerability exposure.
Three vulnerabilities are triggered exclusively by {\toolname}.

\end{abstract}
\maketitle


\section{Introduction}
\label{sec:intro}
Open-source software is widely adopted in software development as third-party libraries (TPLs), which enhance development efficiency~\cite{wu2023ossfp}.
However, it also introduces potential security risks to client programs, as errors or vulnerabilities in TPLs may propagate into dependent client programs~\cite {zhan2021research}.
For instance, the Log4Shell vulnerability impacts millions of client programs~\cite{log4j,hiesgen2022race}.
A mitigation strategy is to update the vulnerable libraries.
However, this process is time-consuming and risky.
Developers must make extensive code changes to adapt to new interfaces~\cite{wang2020empirical, mirhosseini2017can}, while ensuring that the update does not cause conflicts within the software ecosystem~\cite{bogart2016break} nor introduce new vulnerabilities~\cite{dengchainfuzz}.
Since not all vulnerabilities are exploitable from the client, developers often tend to avoid updating TPLs to prevent major changes~\cite{enck2022top, kang2022test}, which exposes many programs to library vulnerabilities~\cite{tang2022towards}.

To encourage developers to adopt mitigation measures (e.g., updating the library), several approaches have been proposed to detect the exploitability of library vulnerabilities from client programs~\cite{decan2018impact, ponta2020detection, chen2024exploiting, kang2022test, foo2019dynamics}.
For instance, reachability-based approaches~\cite{decan2018impact, ponta2020detection} use version or call graph (CG) analysis to determine whether vulnerable functions are invoked.
However, such techniques are prone to false positives, as they cannot verify whether the necessary exploitation conditions are satisfied.
Test-based approaches~\cite{chen2024exploiting, kang2022test, zhou2024magneto} mitigate this issue by generating tests to trigger library vulnerabilities from client programs, thereby evaluating their exploitability.
Nonetheless, these approaches are usually based on proof-of-concepts (PoCs) of vulnerabilities, which are often unavailable~\cite{ruan2024vulzoo, bhuiyan2023secbench, householder2020historical}, limiting their applicability.

\begin{figure*}[t]
\centerline{\includegraphics[width=0.82\linewidth]{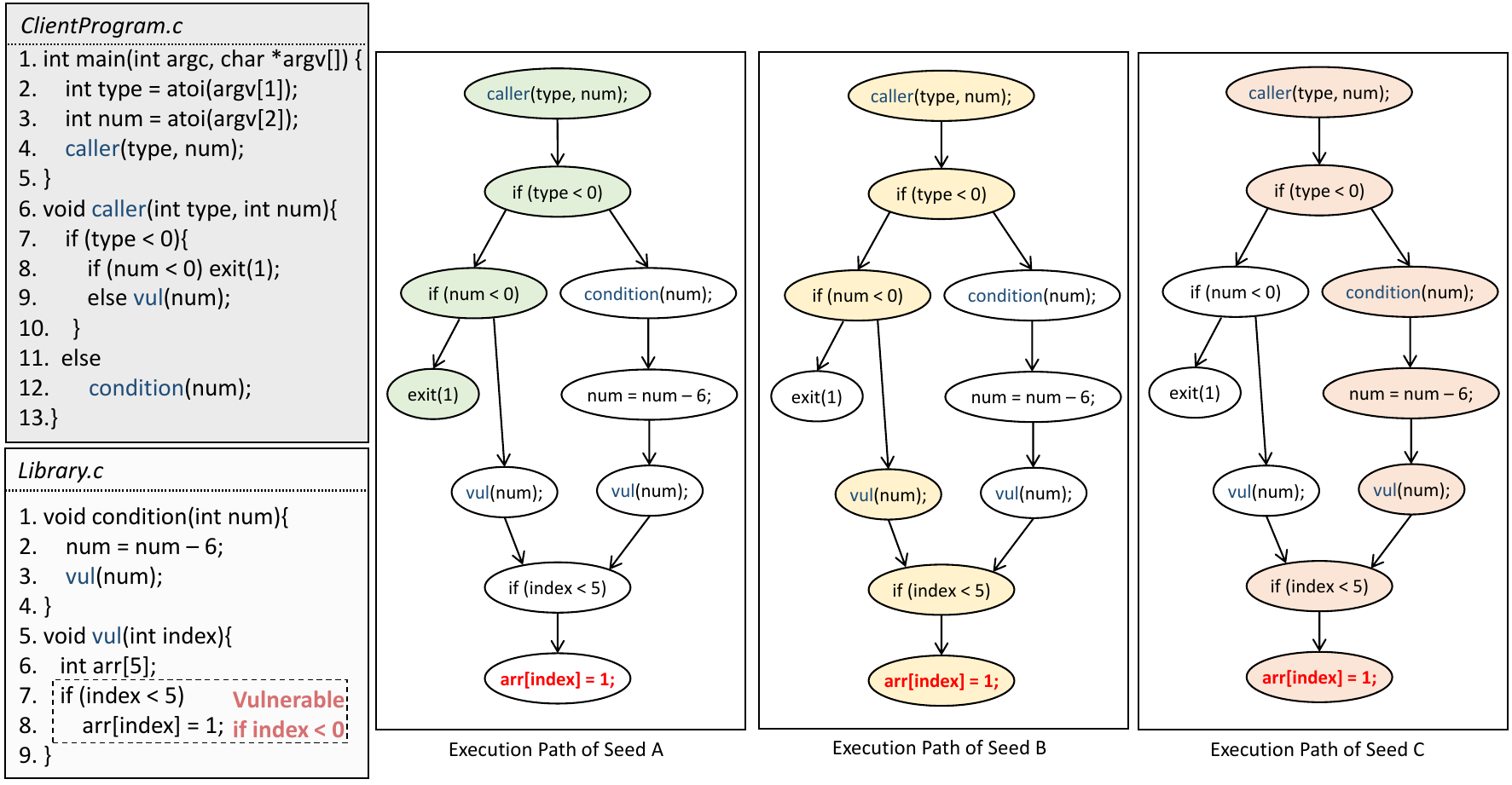}}
\vspace{-0.3cm}
\caption{Example is a library vulnerability that is exploitable with \(type = 1\) and \(num = 5\).}

\label{figlimit}
\vspace{-0.3cm}
\end{figure*}

Given the success of directed greybox fuzzing (DGF) in vulnerability reproduction~\cite{aflgo, li2024sdfuzz}, we extend DGF to trigger library vulnerabilities from client programs without PoCs.
Unlike single-program DGF, this cross-program scenario involves complex inter-layer dependencies connected through dynamic linking~\cite{dengchainfuzz}.
The resulting path explosion and structural boundaries of these cross-program traces introduce two challenges in distance calculation and mutation strategies. 
We illustrate these challenges using the example in Figure~\ref{figlimit}, where three seeds correspond to different execution paths. Triggering the vulnerability requires mutating seed C to generate an input where \(num == 5\).

\noindent \textbf{Challenge~1: The preference for Seeds on Shorter Reachable Paths.}
To enhance the client's security, client developers typically hope to test as many reachable paths as possible.
However, existing DGF approaches prefer selecting seeds that execute along shorter reachable paths~\cite{wen2024empirical, du2022windranger, luo2023selectfuzz}.
This bias is critically exacerbated by the path explosion in cross-program scenarios.
As illustrated in Figure~\ref{figlimit}, seed A fails to reach the target site, while seeds B and C reach it via different execution paths.
Since existing approaches (e.g., AFLGo~\cite{aflgo}) usually calculate distance as the average of basic block distances in the execution path to the target sites, they penalize longer execution paths, yielding rankings such that \(B > A > C\).
This ranking leads fuzzers to prioritize seeds like B and A, while potentially ignoring seed C, resulting in insufficient testing of the reachable path corresponding to seed C.

\noindent \textbf{Challenge~2: Excessive Mutation.}
In cross-program scenarios, inputs must survive the client program's parsing and validation logic before ever reaching the TPL.
However, existing DGF mutation strategies typically select operators without considering the execution stages across program boundaries~\cite{wen2024empirical}.
This often leads to the issue of excessive mutation: coarse-grained operators introduce drastic changes, and the resulting test cases often deviate from the target.
As shown in Figure~\ref{figlimit}, coarse-grained mutations on seed C easily break previously satisfied constraints (e.g., \(type \ge 0\)).
Although Chen et al.~\cite{chen2018hawkeye} adopt more fine-grained mutation operators for the seeds that reach the target sites, the binary classification of seeds based on whether they reached the target site alone cannot adapt to complex real-world scenarios.

In this paper, we propose a novel DGF-based approach to detect \textbf{LI}brary \textbf{V}ulnerabilities' \textbf{E}xploitability from client programs, named \textbf{{\toolname}}.
Specifically, {\toolname} maintains a target tuple \(\langle \texttt{CT}, \texttt{VT} \rangle\) to extend DGF approaches to cross-program scenarios.
For \textbf{Challenge~1}, {\toolname} proposes a novel Abstract Path Mapping mechanism.
Our insight is that the preference stems primarily from the inability of existing evaluation methods to compare seeds on different reachable paths.
For example, in Figure~\ref{figlimit}, the comparison result between seed A and seed B is reasonable, but wrong results are obtained when comparing seed A or seed B with seed C.
{\toolname} addresses this issue by mapping execution paths to unified reachable paths to evaluate seed risks.
For \textbf{Challenge~2}, {\toolname} proposes a risk-based adaptive mutation to generate a customized
mutation strategy for each seed based on its risk.
Unlike existing DGF approaches that focus on optimizing the power scheduling, {\toolname} focuses on the selection of mutation operators, applying more fine-grained operators to high-risk seeds.
This mechanism mitigates the risk of excessive mutation that can arise from the random selection of mutation operators.

To evaluate {\toolname}, we construct a dataset with 61 real-world cases involving 42 vulnerabilities across seven libraries.
We choose AFLGo~\cite{aflgo}, SelectFuzz~\cite{luo2023selectfuzz}, WindRanger~\cite{du2022windranger}, and AFL++~\cite{fioraldi2020afl++} with different configurations as baselines.
The results show that, compared to baselines, {\toolname} discovers 37\%, 40\%, 45\%, 69\%, 195\%, 155\%, and 51\% more target-reachable paths and detects vulnerabilities 5.85x, 4.74x, 6.25x, 7.08x, 5.88x, 4.80x, and 5.52x faster, respectively.
{\toolname} detects the highest number of vulnerabilities, including three exclusively triggered by {\toolname}.

The contributions of our paper are summarized as follows:
\begin{itemize}[leftmargin=*]
    \item We propose a novel Abstract Path Mapping mechanism. By projecting execution paths onto a unified abstract path, the fuzzer compares seeds on different reachable paths, thereby avoiding the preference.
    \item We implement the prototype of {\toolname} to detect the exploitability of library vulnerabilities from client programs without relying on PoCs.
    \item We construct a dataset containing 61 real-world cases. Experimental results indicate that our approach outperforms baselines in detecting the exploitability of library vulnerabilities from client programs.
\end{itemize}

\section{Background}
\label{sec:background}

\subsection{Vulnerability Management in TPLs}

Many approaches~\cite{zhao2023software, decan2018impact} identify vulnerable libraries by analyzing the versions of TPLs, which often introduce numerous false alarms~\cite{ponta2020detection,foo2019dynamics}.
To mitigate this, Ponta et al.~\cite{ponta2020detection} and Foo et al.~\cite{foo2019dynamics} check the reachability of vulnerable functions by analyzing CGs and control flow graphs (CFGs), but they do not consider whether the client program can provide parameters that can trigger the vulnerability to library interface functions~\cite{kang2022test}.
To alleviate this problem, several approaches~\cite{chen2024exploiting,kang2022test, zhou2024magneto} are proposed to generate tests for client public functions to detect the exploitable library Vulnerabilities from client programs.
These approaches~\cite{kang2022test,chen2024exploiting,zhou2024magneto,dengchainfuzz} employ techniques such as genetic algorithms, LLMs, and taint analysis to modify existing PoCs of library vulnerabilities, thereby generating downstream PoCs that trigger library vulnerabilities via client programs.
However, for security reasons, although information about vulnerabilities (e.g., vulnerable functions or potential execution paths) is publicly available~\cite{li2024sdfuzz, nguyen2020binary, simsek2025pocgen}, directly usable PoCs are often unavailable~\cite{ruan2024vulzoo, bhuiyan2023secbench, householder2020historical}.
Therefore, we propose {\toolname} to detect exploitable library vulnerabilities from client programs through DGF, which is widely used in reproducing vulnerabilities.

\begin{figure*}[t]
\centerline{\includegraphics[width=0.79\linewidth]{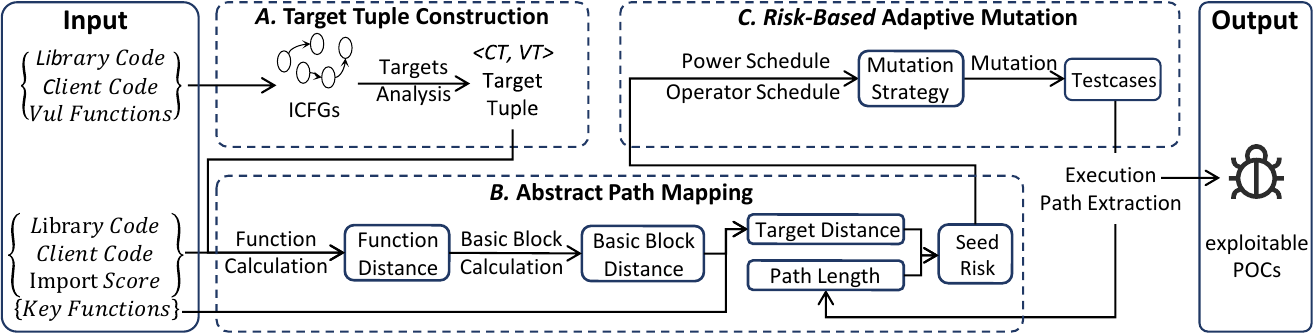}}
\vspace{-0.3cm}
\caption{
The Overall of {\toolname}. 
Module A is Section~\ref{targetanalyzis}, Module B is Section~\ref{distance}, and Module C is Section~\ref{ams}.
}

\label{fig:overall}
\vspace{-0.2cm}
\end{figure*}

\subsection{Fuzzing}
\label{dgf}
Fuzzing~\cite{miller1990empirical} is a widely used technique in software testing~\cite{manes2019art}.
Generally, a fuzzer generates test cases through mutation or other techniques to test programs and saves valuable test cases as seeds for subsequent mutations.
Building on this, B{\"o}hme et al. propose directed greybox fuzzing~\cite{aflgo} (DGF) to focus testing target sites in a program.
Existing approaches~\cite{GF1, GF2, gf3,li2024sdfuzz,nguyen2020binary} analyze the proximity of seeds to target sites (e.g., distance, trajectory similarity, and reachability) by combining vulnerability descriptions and target sites, thereby scheduling power.
For instance, AFLGo~\cite{aflgo}, the most classic DGF method, calculates the shortest distance from each basic block in the seed execution path to the target sites and takes the average of these distances as the seed distance.
AFLGo also employs an annealing-based algorithm, which calculates the power based on seed distance, giving seeds closer to the target sites more opportunities for mutation.
However, existing DGF approaches are often limited by the distance calculation methods and the strategies for selecting mutation operators.


\section{Approach}

Figure~\ref{fig:overall} is the overall of {\toolname}, which comprises three core modules: \textit{Target Tuple Construction}, \textit{Abstract Path Mapping}, and \textit{Risk-Based Adaptive Mutation}.
The Target Tuple Construction module constructs the target tuple to perform all subsequent distance calculations within a single program, thereby extending DGF to the current scenario.
The Abstract Path Mapping mechanism maps execution paths to a unified reachable path by introducing the length of the reachable path, then calculating the relative distance to represent the seed risk (i.e., the potential of a seed to trigger vulnerabilities).
This approach encourages the fuzzer to fairly compare seeds that cause the program to execute along different reachable paths.
The Risk-Based Adaptive Mutation module creates a custom mutation strategy for each seed, allocating more power and applying finer-grained operators to higher-risk seeds.
Similar to existing DGF~\cite{li2024sdfuzz, nguyen2020binary}, {\toolname} takes source code and vulnerability function as inputs.
Moreover, {\toolname} allows testers to provide import scores for different vulnerable functions and key information (e.g., \verb|condition| in Figure~\ref{figlimit}) to enhance its effectiveness.

\subsection{Target Tuple Construction}
\label{targetanalyzis}
The reachable paths to the vulnerable functions always include specific client functions (i.e., caller functions), which are located at the end of the client program’s call chain.
Based on this, unlike prior DGF studies~\cite{aflgo, luo2023selectfuzz, du2022windranger}, 
{\toolname} constructs target tuple \(\langle \texttt{CT}, \texttt{VT} \rangle\) as the target site.
{\toolname} then divides the execution paths into client execution path and library execution path, and specifies \texttt{CT} and \texttt{VT} as targets for them, respectively, which allows all subsequent distance evaluations to occur within a single program's context.
We define Vulnerability Targets and Caller Targets as:

\noindent \textbullet \textbf{Vulnerability Targets (\texttt{VT}).}
\texttt{VT} is defined as the set of the first basic block of each vulnerable function: \( \{ \text{FBB}(f) \mid f \in \mathcal{V} \}\), where \(\text{FBB}(f)\) represents the first basic block of the function \( f \) and \( \mathcal{V} \) is the set of vulnerable functions in TPLs.

\noindent \textbullet \textbf{Caller Targets (\texttt{CT}).}
\texttt{CT} is defined similarly as \(\{ \text{FBB}(f) \mid f \in \mathcal{CC} \}\), where \( \mathcal{CC} \) is the set of caller functions.

To construct this tuple, {\toolname} builds ICFGs for the client program and its library to identify all library functions that can reach a vulnerable function, forming the set A.
{\toolname} then locates all client functions that invoke functions in set A, producing the set \(\mathcal{CC}\).
Meanwhile, the vulnerable functions comprise the set \(\mathcal{V}\).
Finally, {\toolname} parses the compiled binaries to collect the first basic blocks of all functions in set \(\mathcal{V}\) and \(\mathcal{CC}\) to form the target tuple  \(\langle \texttt{CT}, \texttt{VT} \rangle\).
The set \(\mathcal{CC}\) may be incomplete due to limitations in ICFG analysis, which is why we do not merge the call graph directly.
Based on the target tuple, {\toolname} mitigates this issue in the power schedule of the Risk-Based Adaptive Mutation module (Section~\ref{scorec}).
Additionally, {\toolname} uses the first basic blocks instead of line numbers to avoid mismatches caused by the compile environment between the source code and the compiled binary.

\subsection{Abstract Path Mapping.}
\label{distance}
To fairly compare seeds on different reachable paths and mitigate the preference for seeds on shorter paths, {\toolname} utilizes an Abstract Path Mapping mechanism.
Instead of relying on absolute distances, this mechanism evaluates a seed risk (i.e., the potential of a seed to trigger vulnerabilities) based on its relative execution progress.
Specifically, {\toolname} normalizes the distance of the execution path to the target tuple (i.e., target distance) by using the estimated length of its corresponding reachable path (i.e., path length) as an upper bound, thereby evaluating the seed risk.
The target distance is calculated based on the basic block distances, which are derived from function distances.
The path length is defined as the sum of the distance from the entrance function to the current execution path and the target distance.
This formulation is inspired by the \(A^*\) search heuristic \(f(n) = g(n) + h(n)\)~\cite{hart1968formal}, combining the exact traversed cost (\(g\)) with the estimated remaining cost (\(h\)).
Notably, this heuristic serves to normalize relative execution progress across paths, rather than strictly bounding the path length.


\subsubsection{Function Calculation}
\label{fld}
The function distance \(d_{f}(n, T_{f})\) represents the shortest weighted distance between a function \(n\) and a set of target functions \(T_{f}\), where \(T_{f}\) refers to either \(\mathcal{V}\) or \(\mathcal{CC}\).
To prioritize the detection of high-risk vulnerabilities, {\toolname} assigns a weight \(w_f(t_f)\) to each vulnerable function \(t_f \in \mathcal{V}\), which is specified by the tester.
{\toolname} then analyzes the CGs to determine the reachability between vulnerable functions and functions \(t_f \in \mathcal{CC}\).
The weight of each vulnerable function is evenly distributed among all reachable \(t_f \in \mathcal{CC}\).
The weight \(w_f(t_f), t_f \in \mathcal{CC}\) is defined as the average of the weights it receives from all reachable vulnerable functions.
For subsequent calculations, {\toolname} normalizes the weights of functions in \(\mathcal{CC}\) and \(\mathcal{V}\), respectively.
To prioritize high-risk vulnerabilities without completely ignoring low-risk targets, we define the weighted path length \(\widetilde{P}(n,t_{f})\) as the arithmetic mean of the raw distance and the risk-adjusted distance.
Specifically, the raw distance is \(P_{short}(n,t_{f})\), and the risk-adjusted distance scales it by the vulnerability weight as \((1-w_{f}(t_{f}))P_{short}(n,t_{f})\).
Therefore, the final path length is formulated as \(\widetilde{P}(n, t_{f}) = \frac{P_{short}(n,t_{f}) + (1-w_{f}(t_{f}))P_{short}(n,t_{f})}{2}\).
Finally, following standard DGF practices~\cite{aflgo}, the function distance \(d_f(n, T_f)\) is computed as the harmonic mean of the individual weighted path lengths \(\widetilde{P}(n, t_f)\) for all reachable targets in \(T_f\). If no targets are reachable, the distance is undefined.
Note that, when \(n\) is a client function, \(T_{f}~=~\mathcal{CC}\); when \(n\) is a library function, \(T_{f}~=~\mathcal{V}\).


\subsubsection{Basic Block Calculation}
The basic block distance \(d_b(m, T_b)\) evaluates the proximity from a basic block \(m\) to a set of target basic blocks \(T_b\). To evaluate distances within their respective program contexts,  {\toolname} dynamically assigns \(T_b = \mathcal{CT}\) for basic blocks in the client program, and \(T_b = \mathcal{VT}\) for those in the library.
Then, to calculate \(d_b(m, T_b)\), we follow the standard method in AFLGo~\cite{aflgo}, which evaluates only target-reachable basic blocks.
Specifically, if a reachable block \(m\) invokes a function with the function distance, \(d_b(m, T_b)\) is derived by scaling that function's distance by a constant factor.
Otherwise, it is computed recursively by summing the CFG edge counts and adding the distances of valid successors, using a harmonic mean to aggregate multiple branch paths.
The successor \(m_s\) is a basic block that is reachable from the basic block \(m\) and invokes functions with function distances.

\subsubsection{Seed Risk Evaluation}
\label{SD}


Since {\toolname} operates in a cross-program context guided by the target tuple \((\texttt{CT}, \texttt{VT})\), the abstract path mapping must be performed independently for the client and the library. 
Therefore, {\toolname} divides the execution path of a seed \(s\) into two segments: the client execution path (composed of client basic blocks) and the library execution path (composed of library basic blocks).
Similarly, the corresponding estimated reachable path is divided into the client reachable path and the library reachable path.
The client reachable path spans from the client entrance function to \(\texttt{CT}\), while the library reachable path spans from the library interface function to \(\texttt{VT}\).
Let \(d_{s}(s,\texttt{CT})\) and \(d_{s}(s,\texttt{VT})\) be the target distances of the client and library execution path to the target sites \(\texttt{CT}\) and \(\texttt{VT}\).
Let \(d_{r}(s,\texttt{CT})\) and \(d_{r}(s,\texttt{VT})\) represent the path lengths of the corresponding reachable paths.
The seed risk is defined as a tuple \([R_{client}(s,~\texttt{CT}),~R_{library}(s,~\texttt{VT})]\), calculated as:
\begin{equation}
\renewcommand{\arraystretch}{1.8}
\label{ds}
\resizebox{0.7\linewidth}{!}{
$
\begin{bmatrix}
\mathbb{I}_{d_s(s,\texttt{CT}) \neq -1} \frac{d_s(s,\texttt{CT})}{d_r(s,\texttt{CT})} + \mathbb{I}_{d_s(s,\texttt{CT}) = -1}(-1) \\
\mathbb{I}_{d_s(s,\texttt{VT}) \neq -1} \frac{d_s(s,\texttt{VT})}{d_r(s,\texttt{VT})} + \mathbb{I}_{d_s(s,\texttt{VT}) = -1}(-1)
\end{bmatrix}
$
}
\end{equation}
where \(\mathbb{I}_{d_s(s,\texttt{CT}) \neq -1}\) and \(\mathbb{I}_{d_s(s,\texttt{VT}) \neq -1}\) are indicator functions, indicating that when \(d_s\) is -1 (i.e., invalid distance), the corresponding \(R\) is set to -1; Otherwise, the risk is calculated.
The lower value indicates higher seed risk, which is for the convenience of the Power Schedule and Mutation Operator Schedule in Section~\ref{Mutation}.
After obtaining the risk, we apply a normalization to map the non-negative values of \(R_{client}\) and \(R_{library}\) into the [0, 1] range.
Note that invalid risks with a value of -1 remain unchanged.

By default, the target distances \(d_{s}(s,\texttt{CT})\) and \(d_{s}(s,\texttt{VT})\) are calculated by averaging the basic block distances along the client path and library path.
However, in practical scenarios, testers often possess prior knowledge, such as vulnerability descriptions or bug reports. 
To leverage this information and further accelerate vulnerability exposure, {\toolname} introduces an optional enhancement mechanism.
Specifically, testers can provide key functions derived from such prior knowledge.
{\toolname} then calculates an importance score for each key function based on the predefined target weights (\(w_f\)) of its associated vulnerabilities.
Given that a key function may be associated with multiple vulnerability targets, {\toolname} calculates the importance score of a key function \(n\) as \(\sqrt{\sum_{w_{f} \in W(n)} w_{f}^{2}}\).
Here, \(W(n)\) denotes the set of target weights (\(w_f\)) corresponding to the vulnerabilities associated with the key function \(n\).
This definition ensures that both the high-risk (with higher \(w_f\)) and frequency (associated with multiple vulnerabilities) of key functions contribute to a higher importance score, while avoiding score inflation due to redundant occurrences.
To encourage the prioritization of seeds whose execution paths contain such key functions, let \( {\xi(s)} \) and \( {\gamma(s)}\) denote the client execution path and the library execution path of a seed \(s\), respectively.
Let \(IS(s)\) represent the sum of the importance scores of all key functions in the execution path of the seed \(s\).
The final target distances are:
\begin{equation}
\label{dst}
\renewcommand{\arraystretch}{1.5}
\resizebox{0.9\linewidth}{!}{
$
\begin{bmatrix}
d_s(s,\texttt{CT}) 
 \\
d_s(s,\texttt{VT})
\end{bmatrix} = 
\begin{bmatrix}
 \mathbb{I}_{|\xi(s)|>0} \frac{\sum_{m \in \xi(s)} d_b(m, \texttt{CT})}{|\xi(s)|} + \mathbb{I}_{|\xi(s)|=0}(-1)
\\
\mathbb{I}_{|\gamma(s)|> 0} \frac{\sum_{m \in \gamma(s)} d_b(m, \texttt{VT})}{|\gamma(s)| + IS(s)} + \mathbb{I}_{|\gamma(s)|= 0}(-1)
\end{bmatrix}
$
}
\end{equation}
where \(\left |\gamma(s)\right |\) and \(\rm \left |\xi(s)\right |\) represent the numbers of the basic blocks included in \( {\gamma(s)}\) and \( {\xi(s)} \).
Note that when the execution path contains no reachable basic blocks, the distance is set to -1.
In this formulation, \(d_{s}(s,\texttt{CT})\) represents the average distance of all basic blocks within the client execution path.
For \(d_{s}(s,\texttt{VT})\), the aggregated importance score \(IS(s)\) is incorporated into the denominator.
Therefore, the higher the importance score of a key function, the shorter the target distance for the execution path containing this function, guiding the fuzzer toward the vulnerable logic.

In complex programs, accurately determining the reachable path length via static analysis is prohibitively expensive.
As previously discussed, we approximate this length by summing the target distance (\(d_s\)) and the distance from the entrance function to the execution path.
To maintain mathematical consistency with \(d_s\), which calculates the average distance of target-reachable basic blocks to the target, we similarly define the latter component as the average distance of target-reachable blocks from the entrance function.
For each valid block, its distance is simply the number of target-reachable blocks executed before it.
Therefore, let \(|\xi(s)|\) and \(|\gamma(s)|\) denote the total number of target-reachable basic blocks in the client and library execution paths, respectively.
The path lengths \(d_r(s, \mathcal{CT})\) and \(d_r(s, \mathcal{VT})\) are:
\begin{equation}
\renewcommand{\arraystretch}{1.8}
\label{dsm}
\resizebox{0.85\linewidth}{!}{
$
\begin{bmatrix}
d_r(s,\texttt{CT}) 
 \\
d_r(s,\texttt{VT}) 
\end{bmatrix}
=
\begin{bmatrix}
\mathbb{I}_{|\xi(s)|>0} \left(  \frac{|\xi(s)| - 1}{2} + d_s(s,\texttt{CT}) \right) + \mathbb{I}_{|\xi(s)|=0} (-1)
 \\
 \mathbb{I}_{|\gamma(s)|>0} \left(  \frac{|\gamma(s)| - 1}{2}  + d_s(s,\texttt{VT}) \right) + \mathbb{I}_{|\gamma(s)|=0} (-1)
\end{bmatrix}
$
}
\end{equation}
Similarly, when the corresponding basic block is not contained, the reachable path length is set to -1.
Note that the fuzzer categorizes loop executions into logarithmic hit-count buckets.
This mechanism helps discard seeds that merely spin in loops without uncovering new states, thereby mitigating potential path length distortion caused by trivial loop inflation.


\subsection{Risk-Based Adaptive Mutation}
\label{Mutation}
The mutation strategy for each seed in {\toolname} consists of two components: the power schedule and the mutation operator schedule.
Unlike existing DGF, which primarily focuses on optimizing power schedules, {\toolname} focuses on the mutation operator schedule.
Specifically, we adapt the power schedule to our new risk metric, while the operator schedule dynamically adjusts the ratio of coarse- to fine-grained operators to balance broad path exploration with deep logic exploitation.

\subsubsection{Power Schedule}
\label{scorec}

{\toolname} adopts an annealing-based power scheduling algorithm to schedule the power of each seed according to its risk, similar to the algorithm used in AFLGo~\cite{aflgo}.
While AFLGo schedules power based on seed distance, execution time, and a pre-defined exploitation time, 
{\toolname} replaces the seed distance with seed risk and determines the execution time associated with \(\tilde{R}_{library}(s,\textit{VT})\) as the elapsed time since \(\tilde{R}_{library}(s,\textit{VT})\) first became non-zero.
This design prevents premature switching to exploitation and ensures sufficient exploration of library paths.
Ultimately, the power of a seed is the sum of the power \(Power_{client}\) corresponding to \(\tilde{R}_{client}(s, \textit{CT})\) and the power \(Power_{library}\) corresponding to \(\tilde{R}_{library}(s,\textit{VT})\).
However, as noted in Section~\ref{targetanalyzis}, the set \(\mathcal{CC}\) may be incomplete, which can lead to inaccurate calculations of \(\tilde{R}_{client}(s, \textit{CT})\) and, consequently, \(Power_{client}\) for certain seeds.
To address this issue, for any seed with a non-zero \(Power_{library}\), {\toolname} assigns \(Power_{client}\) the maximum value corresponding to the current execution time (i.e., treating its risk \(\tilde{R}_{client}(s, \textit{CT})\) as 0).
This adjustment ensures that, regardless of whether the function invoking the library interface function belongs to \(\mathcal{CC}\), the value of \(Power_{client}\) for the corresponding seed remains consistent.
Our intuition is that if an input satisfies a subsequent condition (e.g., executing a library function), it must also satisfy its prerequisite (e.g., executing the function that invokes the library interface function).
Therefore, in the absence of prior knowledge, the client risk of all seeds whose execution paths include library functions should be identical and should be greater than that of seeds whose execution paths do not involve any library functions.

\subsubsection{Mutation Operator Schedule}
\label{ams}
Mutation operators are categorized as either coarse-grained or fine-grained, depending on the number of bytes they modify.
Coarse-grained operators apply large-scale changes to seeds, which helps the fuzzer escape current path constraints and discover previously unexplored execution paths, thereby improving path diversity.
Typical coarse-grained mutations include inserting or deleting large byte sequences and replacing entire input segments.
In contrast, fine-grained operators make small-scale modifications, enabling focused exploration around the current execution path to expose deep logic flaws or boundary conditions.
Examples include bit flipping, single-byte replacement, and byte increment/decrement operations.
{\toolname} introduces an adaptive mutation operator schedule that progressively shifts the operator selection.
As a seed's risk increases, it dynamically adjusts the usage ratio of different operators applied to that seed.

To implement this, {\toolname} adopts the mutation operator schedule, which dynamically adjusts the proportion of coarse- and fine-grained operators.
{\toolname} defines two operator sets: \verb|FMS|, containing only fine-grained operators, and \verb|HMS|, containing both coarse- and fine-grained operators.
Each seed undergoes multiple mutations, and for each mutation, {\toolname} randomly selects multiple operators from \verb|FMS| or \verb|HMS|.
{\toolname} calculates a ratio for each seed to determine its probability of selecting mutation operators from \verb|FMS|.
This ratio is based on the seed risk and execution time.
Let \( t \) be the running time of the fuzzer and \( t_x \) be a predefined time.
The fine-grained operator ratio \(FR(t,~\tilde{R}_{client}(s,~\texttt{CT}),~\tilde{R}_{library}(s,~\texttt{VT}) )\) is calculated as follows:
\begin{equation}
\renewcommand{\arraystretch}{1.4}
\resizebox{0.9\linewidth}{!}{
$
\begin{cases}
0.25 (1-20^{-t/t_x} ) \cdot (2^{1-\tilde{R}_{client}(s,~\texttt{CT})}-1)
  &  ,~\text{if }\tilde{R}_{library}= -1 ~and~ \tilde{R}_{client} \ge 0 \\
0.25 (1-20^{-t/t_x} ) \cdot 2^{1-\tilde{R}_{library}(s,~\texttt{VT})}
& ,~\text{if }\tilde{R}_{library}\ge 0\\
\end{cases}
$
}
\end{equation}
For seeds without reaching the library (\(\tilde{R}_{library} = -1\)), \(FR\) is constrained within \([0, 0.25]\). Once a seed enters the library (\(\tilde{R}_{library} \ge 0\)), \(FR\) scales into \([0.25, 0.5]\).
This formula uses \((1-20^{-t/t_x}) \) to control the switch between exploration and exploitation and keep the switch consistent with that in the power scheduling algorithm, thereby avoiding local optima.
Based on this, {\toolname} generates a suitable mutation operator usage strategy for each seed based on risk, rather than relying on the coarse-grained binary classification result of whether the target is reached.
For instance, {\toolname} applies more fine-grained operators to seeds that reach the library compared to those that remain within the client program. 

\section{Evaluation}
We compare {\toolname} to answer the following questions:

\phead{RQ1: How good is covering target-reachable paths?}

\phead{RQ2: How well is reproducing vulnerability?}

\phead{RQ3: How does every module contribute to its performance?}

\vspace{-0.1cm}

\subsection{Dataset Collection}
\label{sec:dataset}
\begin{figure}[t]
\centerline{\includegraphics[width=0.92\linewidth]{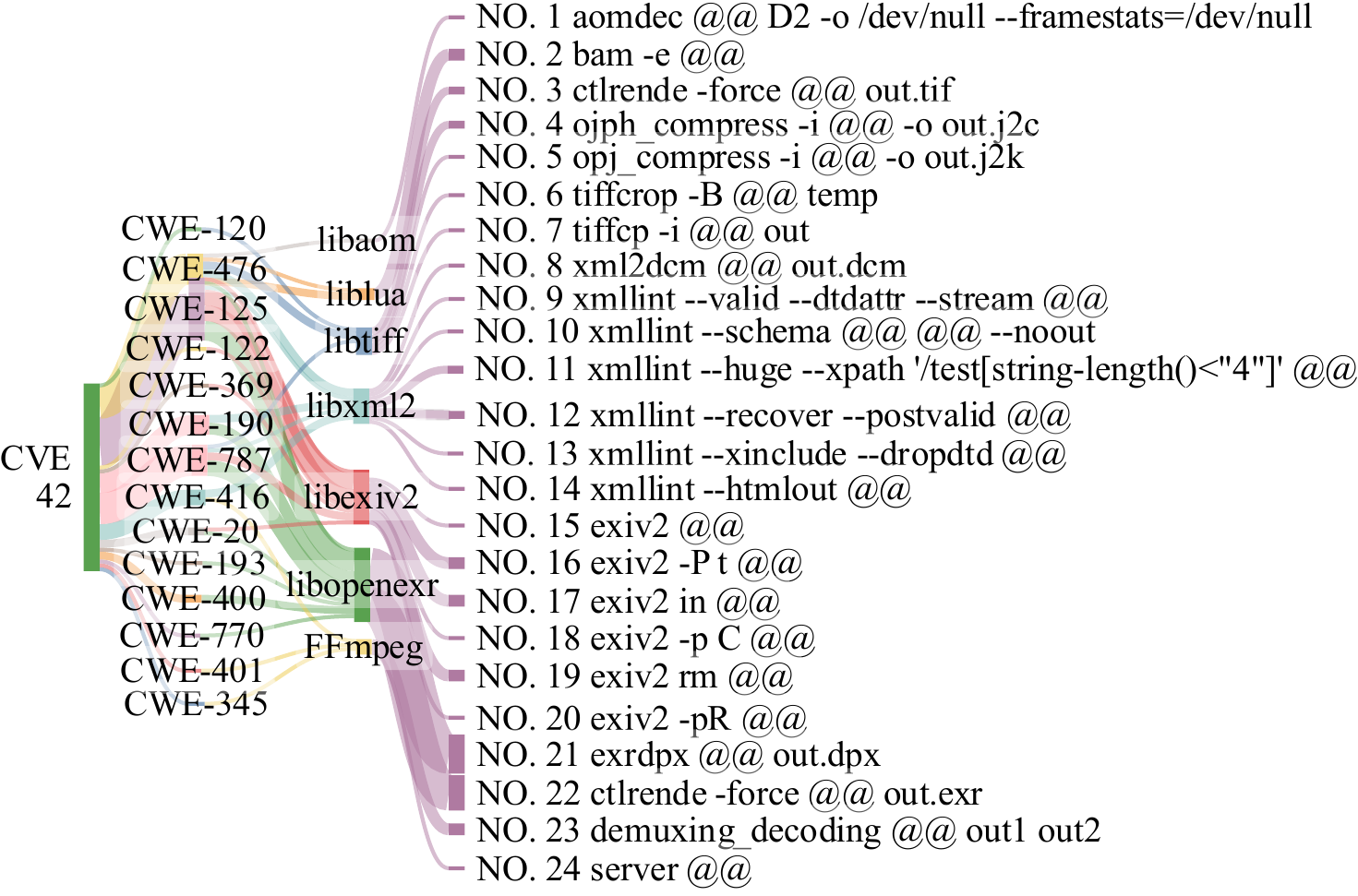}}
\vspace{-0.3cm}
\caption{The Details of the Dataset.}
\vspace{-0.4cm}
\label{fig:dataset}
\end{figure}

To our knowledge, there is currently no complete open-source C/C++ related dataset, even C/C++ client programs face serious security risks~\cite{tang2022towards}.
Although Deng et al.~\cite{dengchainfuzz} conduct experiments in C/C++, their dataset is not completely open source.
Therefore, we collect a new dataset containing 61 real-world cases (i.e., client-library-command-CVE) in which a library vulnerability can be triggered from the client program.
The dataset is presented in Figure~\ref{fig:dataset}, which covers 42 library vulnerabilities across seven libraries, encompassing 14 different vulnerability types.
The highest CVSS score for 42 vulnerabilities is 9.8, while the lowest is 5.3.
Among them, the depth of CVE-2020-15889 is 249, whereas the average depth of the other vulnerabilities is 6.15.
To evaluate the effectiveness of {\toolname} in real scenarios, the dataset follows the following steps to collect exploitable vulnerabilities from real programs.

\subsubsection{Vulnerability collection}
\label{vulcollect}
We collect vulnerabilities from the TPLs of OSS-FUZZ's programs~\cite{serebryany2017oss} as follows:

\noindent \textbf{Step-1}: Program collection. We filter the programs using the following criteria to identify more third-party libraries that are widely used in real-world programs: 
(a) the language of the program is C/C++; 
(b) the program can be built successfully and be fuzzed by AFL or AFL++; 
(c) the source code is available on GitHub.
We collect 181 programs from OSS-FUZZ.

\noindent \textbf{Step-2}: TPL collection. We identify the TPLs using \textsc{ldd}~\cite{ldd} in the collected programs and extract the corresponding Common Platform Enumeration 2.3 (CPE)~\cite{cpe} for each TPL to obtain 174 libraries.
We then select widely used libraries that cover a range of domains, such as audio processing (e.g., FFmpeg), XML parsing (e.g., libxml2), image processing (e.g., libtiff), scripting language parsing (e.g., lua), video encoders (e.g., aomedia), and network protocols (e.g., OpenSSL). 
Consequently, we select 18 TPLs.

\noindent \textbf{Step 3: Vulnerability collection.} 
For each library, we collect vulnerability reports from the National Vulnerability Database (NVD)~\cite{nvd} reported since 2020 based on the CPE.
To ensure that these library vulnerabilities can be triggered within our testing environment, we utilize existing PoCs to reproduce them.
Finally, we collect 133 vulnerabilities and their PoCs across ten libraries.

\begin{table}[]
\caption{The Details of Baselines}
\centering 
\vspace{-0.3cm}
\resizebox{0.85\linewidth}{!}{
\begin{tabular}{@{}cccc@{}}
\toprule
\textbf{Baseline} & \textbf{Fuzzer} & \textbf{Category} & \textbf{Target}       \\ \midrule
AFLGo-CT          & AFLGo~\cite{aflgo}           & Directed          & Caller Targets        \\
AFLGo-VT          & AFLGo           & Directed          & Vulnerability Targets \\
AFLGo-M           & AFLGo           & Directed          & Vulnerability Targets \\
SelectFuzz-CT     & SelectFuzz~\cite{luo2023selectfuzz}     & Directed          & Caller Targets        \\
SelectFuzz-VT     & SelectFuzz      & Directed          & Vulnerability Targets \\
WindRanger        & WindRanger~\cite{du2022windranger}      & Directed          & Vulnerability Targets \&\& Caller Targets \\ \midrule
AFL++       & AFL++~\cite{fioraldi2020afl++}     & Coverage          & -                     \\ \bottomrule
\end{tabular}
}
\label{table:baseline}
\vspace{-0.4cm}
\end{table}

\begin{table*}[]
\centering
\setlength{\abovecaptionskip}{0.05cm} 
\setlength{\belowcaptionskip}{0.1cm}
\caption{The Average Number of Target-Reachable Paths}
\resizebox{0.75\textwidth}{!}{
\begin{tabular}{@{}c|rr|rr|rr|rr|rr|rr|rr|c@{}}
\toprule
\multicolumn{1}{l|}{\multirow{2}{*}{\textbf{No.}}} & \multicolumn{2}{c|}{\textbf{AFLGo-CT}} & \multicolumn{2}{c|}{\textbf{AFLGo-VT}} & \multicolumn{2}{c|}{\textbf{AFLGo-M}} & \multicolumn{2}{c|}{\textbf{WindRanger}} & \multicolumn{2}{c|}{\textbf{SelectFuzz-CT}} & \multicolumn{2}{c|}{\textbf{SelectFuzz-VT}} & \multicolumn{2}{c|}{\textbf{AFL++}} & \multicolumn{1}{c}{\textbf{\toolname}} \\ \cline{2-16} 
\multicolumn{1}{l|}{}                              & $\mu P_{vt}$         & $R_{p}$         & $\mu P_{vt}$         & $R_{p}$         & $\mu P_{vt}$         & $R_{p}$        & $\mu P_{vt}$          & $R_{p}$          & $\mu P_{vt}$            & $R_{p}$           & $\mu P_{vt}$            & $R_{p}$           & $\mu P_{vt}$           & $R_{p}$          & $\mu P_{vt}$                                          \\ \hline
1                                                  & 1066.30              & 1.75*            & 1088.30              & 1.72*            & 1063.80              & 1.76*           & CE                    & CE               & 379.40                  & 4.92*              & 319.10                  & 5.85*              & {\ul 3224.90}          & 0.58*             & 1867.20                                               \\
2                                                  & 2458.30              & 1.31*            & 2574.50              & 1.25*            & 2545.60              & 1.26*           & CE                    & CE               & CE                      & CE                & CE                      & CE                & 521.60                 & 6.15*             & {\ul 3210.40}                                         \\
3                                                  & 285.50               & 1.06            & 244.00               & 1.24*            & 232.00               & 1.30*           & 218.10                & 1.39*             & 58.50                   & 5.17*              & 101.70                  & 2.98*              & 125.20                 & 2.42*             & {\ul 302.60}                                          \\
4                                                  & 63.30                & 1.36            & {\ul 114.10}         & 0.75            & 87.00                & 0.99           & CE                    & CE               & CE                      & CE                & CE                      & CE                & 0.10                   & 858.00*           & 85.80                                                 \\
5                                                  & 20.20                & 4.16*            & 12.20                & 6.89*            & 16.90                & 4.97*           & 18.70                 & 4.49*             & 7.00                    & 12.00*             & 10.50                   & 8.00*              & 0.00                   & -                & {\ul 84.00}                                           \\
6                                                  & 0.00                 & -               & 0.00                 & -               & 0.00                 & -              & CE                    & CE               & 0.00                    & -                 & 0.00                    & -                 & 0.00                   & -                & {\ul 0.00}                                            \\
7                                                  & 3576.50              & 1.22*            & 3675.30              & 1.19*            & 3214.60              & 1.36*           & CE                    & CE               & 1329.00                 & 3.29*              & 1888.30                 & 2.32*              & 3991.30                & 1.10*             & {\ul 4374.50}                                         \\
8                                                  & 5573.80              & 1.17*            & 5794.20              & 1.13*            & 5162.40              & 1.26*           & 5193.50               & 1.26*             & 4281.20                 & 1.53*              & 3103.60                 & 2.10*              & 5490.30                & 1.19*             & {\ul 6530.20}                                         \\
9                                                  & 28.10                & 1.22            & 24.50                & 1.40*            & 25.50                & 1.34*           & 53.80                 & 0.64*             & 23.70                   & 1.44*              & 36.70                   & 0.93              & {\ul 104.70}           & 0.33*             & 34.20                                                 \\
10                                                 & 84.50                & 1.16*            & 87.80                & 1.12*            & 86.70                & 1.13*           & 142.70                & 0.69*             & 81.40                   & 1.21*              & 85.50                   & 1.15*              & {\ul 261.00}           & 0.38*             & 98.10                                                 \\
11                                                 & 5758.60              & 1.15*            & 6029.20              & 1.10*            & 5652.40              & 1.17*           & 4253.50               & 1.56*             & 3484.60                 & 1.90*              & 3939.70                 & 1.68*              & 6336.20                & 1.05*             & {\ul 6621.40}                                         \\
12                                                 & 1445.00              & 1.31*            & 1444.20              & 1.31*            & 1224.60              & 1.54*           & 0.00                  & -                & 1156.20                 & 1.63*              & 1485.80                 & 1.27*              & 0.00                   & -                & {\ul 1889.60}                                         \\
13                                                 & 2193.40              & 1.44*            & 1219.60              & 2.59*            & 1244.00              & 2.54*           & 1461.60               & 2.16*            & 384.00                  & 8.24*              & 493.70                  & 6.41*              & {\ul 4470.10}          & 0.71*             & 3163.90                                               \\
14                                                 & 5796.70              & 1.07*            & 5905.70              & 1.05            & 5579.40              & 1.11*           & 3687.60               & 1.68*             & 3287.70                 & 1.89*              & 3648.10                 & 1.70*              & 5577.60                & 1.11*             & {\ul 6205.20}                                         \\
15                                                 & 21.90                & 1.57*            & 20.90                & 1.64*            & 20.30                & 1.69*           & 6.00                  & 5.72*             & 6.20                    & 5.53*              & 6.70                    & 5.12*              & 19.90                  & 1.72*             & {\ul 34.30}                                           \\
16                                                 & 614.70               & 1.61*            & 595.20               & 1.66*            & 679.70               & 1.45*           & 493.10                & 2.00*             & 546.10                  & 1.81*              & {\ul 1135.90}           & 0.87              & 1056.40                & 0.93             & 987.20                                                \\
17                                                 & 248.40               & 1.96*            & 268.80               & 1.81*            & 263.10               & 1.85*           & 192.50                & 2.52*             & 199.30                  & 2.44*              & 251.00                  & 1.94*              & 282.20                 & 1.72*             & {\ul 486.00}                                          \\
18                                                 & 28.80                & 1.34*            & 29.20                & 1.32*            & 22.50                & 1.72*           & 27.40                 & 1.41*             & 5.00                    & 7.72*              & 5.00                    & 7.72*              & 26.30                  & 1.47*             & {\ul 38.60}                                           \\
19                                                 & 37.20                & 1.63*            & 39.10                & 1.55*            & 35.60                & 1.71*           & 25.60                 & 2.37*             & 11.40                   & 5.32*              & 16.40                   & 3.70*              & 37.30                  & 1.63*             & {\ul 60.70}                                           \\
20                                                 & 203.20               & 1.43*            & 241.70               & 1.21*            & 221.70               & 1.31*           & 225.90                & 1.29*             & 81.80                   & 3.56*              & 92.00                   & 3.17*              & 266.70                 & 1.09             & {\ul 291.30}                                          \\
21                                                 & 1604.30           & 1.18*            & 1434.10              & 1.31*            & 1532.10            & 1.22*           & 1392.50               & 1.35*             & 859.80                  & 2.18*              & 1004.20                 & 1.87*              & {\ul 2191.90}          & 0.86*             & 1875.80                                               \\
22                                                 & 1361.40              & 1.29*            & 1473.60              & 1.19*            & 1365.50              & 1.28*           & 1339.70               & 1.31*             & 772.80                  & 2.27*              & 851.60                  & 2.06*              & {\ul 4242.00}          & 0.41*             & 1754.20                                               \\
23                                                 & 5408.00              & 1.01            & 5401.50              & 1.01            & {\ul 5787.30}        & 0.94           & CE                    & CE               & 4915.50                 & 1.11*              & 1978.80                 & 2.76*              & CE                     & CE               & 5460.70                                               \\
24                                                 & 5319.30              & 1.00            & 4771.10              & 1.12*            & 4939.00              & 1.08*           & CE                    & CE               & CE                      & CE                & CE                      & CE                & CE                     & CE               &  {\ul 5335.10 }                                              \\ \hline
Avg.                                               &                      & 1.37            &                      & 1.40            &                      & 1.45           &                       & 1.69             &                         & 2.95              &                         & 2.55              &                        & 1.51             &                                                       \\ \hline
\end{tabular}
}
\begin{tablenotes}    
    \footnotesize               
    \item The $\mu P_{vt}$ is the mean of \(P_{vt}\) and Avg. is the geometric mean of $R_{p}$. ``CE'' indicates that the fuzzer fail to deploy in the program. The best result is underlined. * shows statistical significance at $p \le 0.05$.
\end{tablenotes}
\vspace{-0.3cm}
\label{table:rq1}
\end{table*}

\subsubsection{Client programs collection}
To collect client programs, we identify potential, real-world programs for each library vulnerability.
Based on the CGs of libraries, we first extracted the library interface functions that may trigger the vulnerability.
Then, we search for potential client programs that invoke these interface functions and have more than 100 stars on GitHub via GitHub's code search API~\cite{api}.
We verify their actual invocation through execution.
We also retain client programs with more than 1,000 lines of code provided by the library repository.
Finally, we gain 45 programs as client programs to try to exploit the library vulnerabilities.

\subsubsection{Library Vulnerability Exploitability Analysis}
To verify whether library vulnerabilities can be exploited from client programs, we attempt to trigger them using the PoCs.
We confirm a successful trigger by matching the stack traces output by ASan and UBSan with the relevant vulnerability descriptions.
Relying on existing PoCs for this verification effectively eliminates the bias and errors associated with manual analysis.
Importantly, while these PoCs establish an objective ground truth for our evaluation, they are never provided as inputs to {\toolname}.


\subsection{Implementation and Experiment Setup}
\noindent\textbf{Implementation.}
we implement a prototype of {\toolname} based on AFLGo~\cite{aflgo}.
We use SVF~\cite{sui2016svf, sui2014detecting} to construct ICFGs, and instrument basic block distance using an LLVM Pass.

\noindent\textbf{Baselines.}
To our knowledge, no existing research detects the exploitability of library vulnerabilities from client programs without PoCs. 
Since the most similar work is DGF, which is used to detect library vulnerabilities through drivers, we select AFLGo~\cite{aflgo}, SelectFuzz~\cite{luo2023selectfuzz}, and WindRanger~\cite{du2022windranger} as our baselines. 
For these DGF tools, we deploy two configurations, setting \texttt{CT}s and \texttt{VT}s as their target sites respectively. 
We also implement AFLGo-M to investigate the feasibility of analyzing the cross-program execution as a whole by extending AFLGo to merge the client and library CGs into a unified ICFG.
Additionally, we include AFL++~\cite{fioraldi2020afl++} as a baseline representing state-of-the-art greybox fuzzing. 
Baseline details are provided in Table~\ref{table:baseline}. 
Some related works, such as Hawkeye~\cite{chen2018hawkeye} and SDFUZZ~\cite{li2024sdfuzz}, are not included because they are not open-sourced at the time of writing.
Beacon~\cite{huang2022beacon} and DAFL~\cite{kim2023dafl} cannot be used in multi-target scenarios.
Titan~\cite{huang2023titan} is also excluded from our baselines because it is designed for single-program scenarios.

\noindent\textbf{Metric.}
To answer RQ1, we use the number of target-reachable paths \(P_{vt}\) to evaluate the capability of covering target-reachable paths~\cite{zhang2023predecessor}.
To answer RQ2, we evaluate the vulnerability reproducing capability by Time-to-Expose (TTE) with Runs (Rs). 
TTE is the time (seconds) required to trigger a known vulnerability, and Rs is the number of runs the vulnerability is successfully triggered, which are commonly used metrics in DGF.
To assess statistical significance, we conduct the Mann-Whitney U Test~\cite{mcknight2010mann} on \(P_{vt}\) and TTE, setting the TTE of unexposed vulnerabilities to 24 hours~\cite{aflgo}.

\noindent\textbf{Experimental Setup.}
All experiments are conducted on an Intel Xeon Gold 6226R CPU (2.90GHz) with 336GB RAM.
Following standard practices~\cite{huang2023titan}, we perform 10 independent 24-hour runs per experiment.
All fuzzers share identical, PoC-free initial seeds sourced from target libraries or public corpora~\cite{seed}.
To verify vulnerabilities, generated seeds are re-executed against ASan/UBSan-instrumented binaries.
Reflecting real-world conditions, {\toolname} and baselines employ a multi-target approach, simultaneously targeting all vulnerabilities reachable via a single client-library command.
Different from other fuzzing experiments, we run {\toolname} and baselines in a cross-program scenario.
Furthermore, the target weights are assigned based on the vulnerabilities' CVSS scores, while functions mentioned in their descriptions are designated as key functions.
Note that, in RQ3, we perform separate ablations on key functions to simulate scenarios where the descriptions are absent.

\subsection{RQ1: Paths Covering Capability}
\label{subsec:rq1}
In this section, a case denotes a single cross-program execution (client-library-command) encompassing multiple CVEs.
We define the improvement ratio \(R_P\) as the \(P_{vt}\) found by {\toolname} divided by that of the baseline per case.
To count \(P_{vt}\), we re-execute generated seeds and verify \texttt{VT} reachability by instrumenting vulnerable functions with trace logs.
Additionally, we aggregate the overall improvement across all cases using the geometric mean of \(R_P\).

The results are shown in Table~\ref{table:rq1}.
Overall, {\toolname} discovers 37\%, 40\%, 45\%, 69\%, 195\%, 155\%, and 51\% more target-reachable paths than AFLGo-CT, AFLGo-VT, AFLGo-M, WindRanger, SelectFuzz-CT, SelectFuzz-VT, and AFL++, respectively.
We conduct the Mann-Whitney U test across the ten runs, confirming that {\toolname} achieves statistically significant improvements (\(p \le 0.05\)) in the vast majority of cases.
These show that {\toolname} has better capability of covering target-reachable paths.
Additionally, in case 6, neither {\toolname} nor other baselines could reach the target site within 24 hours. 
Upon analyzing issue reports and the libtiff~\cite{libtiff} source code, we find that triggering this vulnerability requires the file to use LogLuv encoding, which is essential to reach the target site, but no initial seed satisfies it.
Consequently, the fuzzer requires significant time to produce files with the necessary encoding through random mutations, which makes it difficult for the fuzzer to generate test cases close to the target sites effectively.
In the few cases where a baseline reports a higher \(\mu P_{vt}\) (e.g., AFLGo-VT in case 4, SelectFuzz-VT in case 16, and AFLGo-M in case 23), the differences are not statistically significant (\(p > 0.05\)), which is highly dependent on the randomness of mutations.
For instance, in case 4, we observe that the average number of target-reachable paths in five runs is only 36.8, whereas the other five runs have an average of 191.4 paths.
Additionally, in some cases, {\toolname} does not outperform AFL++.
AFL++ integrates multiple mutation strategies and enables deterministic mutation, which mutates inputs regularly, systematically modifies each byte or bit, and tries to cover all possible conditions and boundaries, thereby exploring more paths.
Although these operators can improve coverage and demonstrate better target-reachable paths coverage when the time budget is sufficient, they introduce additional time overhead (e.g., inverting each byte in turn takes a lot of time) and fail to accelerate the triggering of specific vulnerabilities effectively.
The experimental results in Section~\ref{subsec:rq2} also support this conclusion.

\greyboxb{Summary for RQ1:} {
In most cases, {\toolname} discovers more target-reachable paths compared to baselines, demonstrating an average improvement of 37\%, 40\%, 45\%, 69\%, 195\%, 155\%, and 51\%, which indicates that {\toolname} has a better capability of covering target-reachable paths.
}

\begin{table*}[]
\centering
\renewcommand{\arraystretch}{1.3} 
\caption{Time-to-Expose Results from Dataset. }
\vspace{-0.3cm}
\resizebox{0.89\textwidth}{!}{
\setlength{\tabcolsep}{3pt} 
\begin{tabular}{@{}lr|rrr|rrr|rrr|rrr|rrr|rrr|rrr|rr@{}}
\toprule
\multicolumn{1}{c|}{\multirow{2}{*}{\textbf{CVE}}} & \multirow{2}{*}{\textbf{No.}} & \multicolumn{3}{c|}{\textbf{AFLGo-CT}}                & \multicolumn{3}{c|}{\textbf{AFLGo-VT}}                      & \multicolumn{3}{c|}{\textbf{AFLGo-M}}                       & \multicolumn{3}{c|}{\textbf{WindRanger}}              & \multicolumn{3}{c|}{\textbf{SelectFuzz-CT}}           & \multicolumn{3}{c|}{\textbf{SelectFuzz-VT}}                 & \multicolumn{3}{c|}{\textbf{AFL++}}                  & \multicolumn{2}{c}{\textbf{\toolname}} \\ \cline{3-25} 
\multicolumn{1}{c|}{}                              &                               & Rs & $\mu TTE$ (s)   & $R_{tte}$ & Rs & $\mu TTE$ (s)         & $R_{tte}$ & Rs & $\mu TTE$ (s)         & $R_{tte}$& Rs & $\mu TTE$ (s)   & $R_{tte}$ & Rs & $\mu TTE$ (s)   & $R_{tte}$& Rs & $\mu TTE$ (s)         & $R_{tte}$ & Rs & $\mu TTE$ (s)        & $R_{tte}$ & Rs                 & $\mu TTE$ (s)                  \\ \hline
\multicolumn{1}{l|}{2020-36130}                     & 1                              & 10   & 125.01        & 3.49*                   & 10   & 206.37        & 5.77*                   & 10   & 213.48         & 5.97*                   & CE   & CE            & CE                          & 10   & 70.28         & 1.96*                                                & 10   & 104.73         & 2.93*                   & 10   & 2229.53       & 62.31*                   & 10                   & {\ul 35.78}                    \\
\multicolumn{1}{l|}{2023-2908}                      & 3                              & 10   & 12288.40      & 6.52*                   & 10   & 7667.62       & 4.07*                   & 10   & 21830.38       & 11.58*                   & 8    & 17732.00      & 9.40*                   & 0    & TO            & -                                                        & 10   & 6779.31        & 3.60*                   & 8    & 10200.24      & 5.41*                   & 10                   & {\ul 1885.48}                  \\
\multicolumn{1}{l|}{2023-2908}                      & 4                              & 6    & 36582.14      & 10.53*                   & 10   & 20496.89      & 5.90*                   & 10   & 11724.23       & 3.37*                   & 0    & TO            & -                          & 0    & TO            & -                                                       & 0    & TO             & -                            & 0    & TO            & -                          & 10                   & {\ul 3474.33}                  \\
\multicolumn{1}{l|}{2023-2908}                      & 5                              & 10   & 40969.94      & 24.51*                   & 7    & 35280.16      & 21.10*                   & 8    & 48633.27       & 29.09*                   & 10   & 6779.87       & 4.06*                   & 0    & TO            & -                                                       & 8    & 29335.90       & 17.55*                   & 0    & TO            & -                          & 10                   & {\ul 1671.75}                  \\
\multicolumn{1}{l|}{2021-3537}                      & 12                             & 5    & 31646.85      & 3.10*                   & 6    & 49965.42      & 4.89*                   & 3    & 38588.55       & 3.77*                   & 0    & TO            & -                          & 1    & 81258.04      & 7.95*                                                & 10   & {\ul 10105.33} & 0.99                   & 0    & TO            & -                          & 10                   & 10223.66                       \\
\multicolumn{1}{l|}{2021-3517}                      & 12                             & 9    & 441.97        & 6.79*                   & 10   & 298.71        & 4.59*                   & 10   & 2195.89        & 33.72*                   & 0    & TO            & -                          & 10   & 176.18        & 2.71*                                                & 10   & 212.99         & 3.27*                   & 0    & TO            & -                          & 10                   & {\ul 65.13}                    \\
\multicolumn{1}{l|}{2021-3518}                      & 13                             & 8    & 7995.90       & 29.92*                   & 10   & 15638.62      & 58.52*                   & 7    & 11365.94       & 42.54*                   & 10   & 564.35        & 2.11*                   & 5    & 23363.90      & 87.44*                                                & 4    & 25730.71       & 96.29*                   & 10   & 1135.36       & 4.25*                   & 10                   & {\ul 267.21}                   \\
\multicolumn{1}{l|}{2020-24977}                     & 14                             & 0    & TO            & -                          & 0    & TO            & -                          & 0    & TO             & -                         & 0    & TO            & -                          & 0    & TO            & -                                                       & 0    & TO             & -                         & 0    & TO            & -                          & 2                    & {\ul 17100.59}                 \\
\multicolumn{1}{l|}{2021-37616}                     & 16                             & 10   & 5264.42       & 8.34*                   & 10   & 2558.37       & 4.05*                   & 10   & 6349.41        & 10.06*                   & 0    & TO            & -                          & 10   & 4549.62       & 7.21*                                                & 10   & 6063.60        & 9.61*                   & 10   & 18959.47      & 30.04*                   & 10                   & {\ul 631.19}                   \\
\multicolumn{1}{l|}{2021-34335}                     & 16                             & 5    & 3838.94       & 2.62*                   & 9    & 9607.61       & 6.56*                   & 9    & 8468.95        & 5.78*                   & 0    & TO            & -                          & 8    & 8088.08       & 5.52*                                                & 10   & 10699.58       & 7.31*                   & 9    & 28358.34      & 19.36*                   & 10                   & {\ul 1464.68}                  \\
\multicolumn{1}{l|}{2021-29457}                     & 17                             & 10   & 4204.49       & 10.34*                   & 10   & 2387.48       & 5.87*                   & 10   & 2770.45        & 6.81*                   & 8    & 34129.63      & 83.90*                   & 10   & 2521.32       & 6.20*                                                & 10   & 1736.79        & 4.27*                   & 10   & 9932.46       & 24.42*                   & 10                   & {\ul 406.78}                   \\
\multicolumn{1}{l|}{2021-29458}                     & 17                             & 10   & 14678.23      & 11.06*                   & 10   & 7234.35       & 5.45*                   & 10   & 4892.60        & 3.69*                   & 10   & 14341.55      & 10.80*                   & 10   & 5776.37       & 4.35*                                                & 10   & 4319.27        & 3.25*                   & 10   & 6198.49       & 4.67*                   & 10                   & {\ul 1327.31}                  \\
\multicolumn{1}{l|}{2021-29470}                     & 17                             & 10   & 4248.12       & 7.87*                   & 10   & 4371.51       & 8.10*                   & 10   & 3507.01        & 6.50*                   & 9    & 2834.14       & 5.25*                   & 9    & 12438.65      & 23.06*                                                & 8    & 8785.42        & 16.28*                   & 10   & 9612.30       & 17.82*                   & 10                   & {\ul 539.50}                   \\
\multicolumn{1}{l|}{2021-37618}                     & 18                             & 10   & 197.11        & 8.13*                   & 10   & 174.74        & 7.21*                   & 10   & 124.69         & 5.14*                   & 10   & 1084.19       & 44.71*                   & 10   & 281.96        & 11.63*                                                & 10   & 146.77         & 6.05*                   & 10   & 119.22        & 4.92*                   & 10                   & {\ul 24.25}                    \\
\multicolumn{1}{l|}{2021-37619}                     & 19                             & 10   & 11830.93      & 16.72*                   & 10   & 4433.38       & 6.27*                   & 10   & 2720.99        & 3.85*                   & 10   & 4658.96       & 6.59*                   & 4    & 72428.38      & 102.38*                                               & 9    & 50446.41       & 71.31*                   & 10   & 4276.43       & 6.04*                   & 10                   & {\ul 707.45}                   \\
\multicolumn{1}{l|}{2021-29457}                     & 19                             & 10   & 105.14        & 4.35*                   & 10   & 132.66        & 5.49*                   & 10   & 122.99         & 5.09*                   & 0    & TO            & -                          & 10   & 579.68        & 23.98*                                                & 10   & 160.91         & 6.66                   & 10   & 3325.55       & 137.58*                  & 10                   & {\ul 24.17}                    \\
\multicolumn{1}{l|}{2020-18831}                     & 20                             & 10   & 2598.13       & 46.07*                   & 10   & 2134.72       & 37.85*                   & 10   & 2606.90        & 46.22*                   & 0    & TO            & -                          & 10   & 1105.05       & 19.59*                                                & 10   & 237.05         & 4.20*                   & 10   & 786.66        & 13.95*                   & 10                   & {\ul 56.40}                    \\
\multicolumn{1}{l|}{2021-3478}                      & 21                             & 10   & 2993.36       & 2.86*                   & 10   & 3071.03       & 2.93*                   & 10   & 2260.38        & 2.86                    & 10   & 3046.95       & 2.91*                   & 10   & 1821.44       & 1.74*                                                & 10   & 1812.78        & 1.73*                   & 10   & {\ul 215.73}  & 0.21*                   & 10                   & 1048.43                        \\
\multicolumn{1}{l|}{2021-20302}                     & 21                             & 1    & 83330.53      & 2.51*                   & 3    & {\ul 8283.01}        & 0.10                   & 3    & 23145.20       & 2.07*                   & 0    & TO            & -                          & 2    & 45453.34      & 1.37*                                                & 2    & 79726.99       & 2.40*                   & 0    & TO            & -                          & 8                    & 33187.90                       \\
\multicolumn{1}{l|}{2020-11764}                     & 21                             & 3    & 64553.74      & 2.07*                   & 1    & 37558.00      & 1.20*                   & 1    & 54120.88       & 2.04*                   & 0    & TO            & -                          & 2    & 64328.93      & 2.06*                                                & 1    & 50374.13       & 1.61*                   & 2    & 40483.06      & 1.30*                   & 6                    & {\ul 31226.66}                 \\
\multicolumn{1}{l|}{2020-11765}                     & 21                             & 2    & 17058.01      & 1.29*                   & 2    & 35330.87      & 2.67*                   & 5    & 34402.79       & 2.97*                   & 4    & 55210.72      & 4.17*                   & 2    & 52167.23      & 3.94*                                                & 2    & 31964.37       & 2.42*                   & 6    & 33677.27      & 2.55*                   & 7                    & {\ul 13227.39}                 \\
\multicolumn{1}{l|}{2020-16587}                     & 21                             & 6    & 27993.75      & 2.04*                   & 2    & 18304.84      & 1.33*                   & 1    & 60019.13       & 4.37*                   & 0    & TO            & -                          & 7    & 45913.63      & 3.34*                                                & 3    & 39338.74       & 2.86*                   & 9    & 25873.31      & 1.88                 & 9                    & {\ul 13732.24}                 \\
\multicolumn{1}{l|}{2021-3474}                      & 21                             & 10   & 2990.91       & 2.97*                   & 10   & 2857.82       & 2.84*                   & 10   & 2619.47        & 2.60              & 10   & 2898.11       & 2.88*                   & 10   & 1807.19       & 1.80*                                                & 10   & 1718.90        & 1.71*                   & 10   & 2885.74       & 2.87*                   & 10                   & {\ul 1006.05}                  \\
\multicolumn{1}{l|}{2021-23215}                     & 22                             & 0    & TO            & -                          & 0    & TO            & -                          & 0    & TO             & -                         & 0    & TO            & -                          & 0    & TO            & -                                                       & 0    & TO             & -                         & 0    & TO            & -                          & 3                    & {\ul 23641.52}                 \\
\multicolumn{1}{l|}{2021-23169}                     & 22                             & 9    & 31933.47      & 17.35*                   & 9    & 26141.88      & 14.21*                   & 9    & 54436.44       & 29.58*                   & 4    & 36354.33      & 19.76*                   & 9    & 33137.33      & 18.01*                                                & 9    & 28273.02       & 15.36*                   & 8    & 11958.54      & 6.50*                   & 10                   & {\ul 1840.21}                  \\
\multicolumn{1}{l|}{2021-3478}                      & 22                             & 10   & 732.25        & 6.42*                   & 10   & 331.37        & 2.90*                   & 10   & 936.48         & 8.21*                   & 10   & 357.88        & 3.14*                   & 10   & 248.17        & 2.18*                                                & 10   & 222.19         & 1.95*                   & 10   & 178.25        & 1.56*                   & 10                   & {\ul 114.08}                   \\
\multicolumn{1}{l|}{2021-20302}                     & 22                             & 4    & 52597.81      & 3.08*                   & 5    & 49781.32      & 2.92*                   & 5    & {\ul 15895.76} & 0.93*                   & 0    & TO            & -                          & 5    & 59654.62      & 3.50*                        & 6    & 45199.05       & 2.65*                   & 2    & 46192.27      & 2.71*                   & 10                   & 17053.73                       \\
\multicolumn{1}{l|}{2020-11765}                     & 22                             & 5    & 18214.90      & 4.37*                   & 8    & 24055.42      & 5.77*                   & 5    & 42040.83       & 10.09*                   & 6    & 26221.39      & 6.29*                   & 3    & 36643.15      & 8.80*                                                & 6    & 41603.01       & 9.99*                   & 8    & 32003.48      & 7.68*                  & 8                    & {\ul 4165.56}                  \\
\multicolumn{1}{l|}{2020-16587}                     & 22                             & 4    & 53921.48      & 2.08*                   & 6    & 43080.77      & 1.66*                   & 2    & 68278.78       & 2.63*                   & 0    & TO            & -                          & 8    & 30351.72      & 1.17      & 8    & 40211.29       & 1.55                     & 8    & {\ul 8863.94} & 0.34                   & 9                    & 25941.72                       \\
\multicolumn{1}{l|}{2021-3474}                      & 22                             & 10   & 533.52        & 4.18*                   & 10   & 407.00        & 3.19*                   & 10   & 613.54         & 4.81*                   & 10   & 465.43        & 3.65*                   & 10   & 266.81        & 2.09*                                                & 10   & 289.75         & 2.27*                   & 9    & 579.58        & 4.55*                   & 10                   & {\ul 127.51}                   \\
\multicolumn{1}{l|}{2020-11760}                     & 22                             & 0    & TO            & -                          & 0    & TO            & -                          & 0    & TO             & -                         & 0    & TO            & -                          & 0    & TO            & -                                                       & 0    & TO             & -                         & 0    & TO            & -                          & 1                    & {\ul 1806.09}                  \\ \hline
\multicolumn{2}{c|}{Avg.}                                                            & \multicolumn{3}{c|}{5.85}                                  & \multicolumn{3}{c|}{4.74}                                  & \multicolumn{3}{c|}{6.25}                                   & \multicolumn{3}{c|}{7.08}                                  & \multicolumn{3}{c|}{5.88}                                                               & \multicolumn{3}{c|}{4.80}                                   & \multicolumn{3}{c|}{5.52}                                  & \multicolumn{2}{c}{}                                  \\
\multicolumn{2}{c|}{Rs Sum}                                                        & \multicolumn{3}{c|}{217}                                   & \multicolumn{3}{c|}{228}                                   & \multicolumn{3}{c|}{218}                                    & \multicolumn{3}{c|}{129}                                   & \multicolumn{3}{c|}{185}                                                                & \multicolumn{3}{c|}{216}                                    & \multicolumn{3}{c|}{199}                                   & \multicolumn{2}{c}{273}                               \\ \hline
\end{tabular}
}
\begin{tablenotes}    
    \footnotesize               
    \item The $\mu TTE$ indicates the average $TTE$ (ignore timeout). TO indicates that the vulnerability is not exposed within 24 hours. $R_{tte}$ is the ratio of $\mu TTE$ used by the baseline to $\mu TTE$ used by {\toolname}. Avg. is the geometric mean of $R_{tte}$. Rs Sum is the sum of Rs. The best result of a case is underlined. * shows statistical significance at $p \le 0.05$. 
\end{tablenotes}
\vspace{-0.2cm}
\label{table:rq2}
\end{table*}

\subsection{RQ2: Vulnerability Reproducing Capability}
\label{subsec:rq2}

Table~\ref{table:rq2} shows vulnerability reproduction results within 24 hours.
Vulnerabilities that have not been triggered within 24 hours are not displayed.
Overall, {\toolname} achieves average speedups of 5.85x, 4.74x, 6.25x, 7.08x, 5.88x, 4.80x, and 5.52x compared to baselines.
For most targets, {\toolname} shows a significant improvement (\(p \le 0.05\)) and exhibits superior stability.
Specifically, in total, {\toolname} triggers 56 more vulnerabilities than AFLGo-CT, 45 more than AFLGo-VT, 56 more than AFLGo-M, 144 more than WindRanger, 88 more than SelectFuzz-CT, 57 more than SelectFuzz-VT, and 74 more than AFL++.
Notably, three vulnerabilities (CVE-2020-24977 of case 14, CVE-2021-23215 of case 22, and CVE-2020-11760 of case 22) are triggered solely by {\toolname}. 
Additionally, it is worth noting that AFLGo-M is less effective than AFLGo-VT.
The key reason is that merging call graphs introduces more basic blocks with a larger distance.
These distant blocks tend to dominate the fuzzer's guidance and reduce the influence of those closer to the target.
For instance, if two seeds diverge only at the last basic block (one being closer to the target and the other starting to move away), their seed distances may not differ significantly. As a result, the guidance becomes less effective, leading to greater performance variance in repeated fuzzing runs.
This is quantitatively reflected in the coefficient of variation~\cite{abdi2010coefficient}, which is significantly higher for AFLGo-M than for AFLGo-VT.
In contrast, {\toolname} adopts a more resilient target tuple, which better tolerates incomplete analysis results. It also avoids assigning excessive distances to basic blocks. The geometric mean of its coefficient of variation is lower than that of all baselines, demonstrating its stability.
We investigate the cases where {\toolname} performed poorly on \(\mu TTE\).
In CVE-2021-3537 of case 12, {\toolname} takes more time to reproduce the vulnerability than SelectFuzz-VT, but in another vulnerability (CVE-2021-3517) of case 12, {\toolname} is 2.71x faster than the best-performing baseline, which reflects our focus on high-risk vulnerabilities.
The CVSS score of CVE-2021-3537 is 5.9, while the one of CVE-2021-3517 is 8.6.
Therefore, {\toolname} prioritizes the latter, increasing the time required to reproduce the former.
In CVE-2021-3478 of case 21, AFL++ shows better performance compared to {\toolname}. 
By further analyzing the details, we find that among all triggered vulnerabilities in case 21, this vulnerability has the shallowest depth, which indicates that the fuzzer does not need too much distance information to trigger it.
Therefore, AFL++ can quickly trigger this vulnerability as a GF tool.
In the remaining poorly performing cases (CVE-2021-20302 of case 21, CVE-2021-20302 of case 22, and CVE-2020-16587 of case 22), the \(\mu TTE\) of {\toolname} is not the shortest because the effect of timeout is ignored when calculating \(\mu TTE\).

\begin{figure}[t]
\centerline{\includegraphics[width=0.77\linewidth]{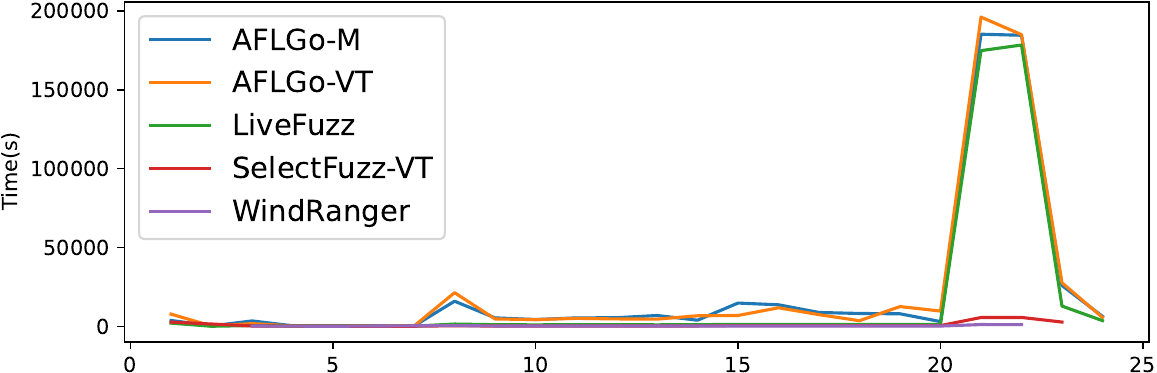}}
\vspace{-0.3cm}
\caption{The Static Analysis Time (s) for Different Cases.}
\label{fig:statitime}
\vspace{-0.3cm}
\end{figure}
To evaluate the performance of {\toolname} more thoroughly, we analyze CWE types and depths of vulnerabilities triggered within 24 hours.
Within 24 hours, {\toolname} successfully triggered 11 types of vulnerabilities.
For buffer overflows (e.g., CWE-787, 122, 125), it achieved a 5.02x speedup over the best baseline (AFLGo-VT) and exclusively exposed two vulnerabilities.
For CWE-476 vulnerabilities, {\toolname} has improved by 4.83x compared with the best-performing baseline.
Additionally, some special vulnerability types, such as CWE-416, {\toolname} achieves a 2.11x improvement over the best-performing baseline.
These results indicate that {\toolname} can detect various vulnerability types and outperforms baselines in most cases.
The average depth of vulnerabilities triggered within 24 hours is 5.42, exceeding that of related studies~\cite{chen2024exploiting, zhou2024magneto}.

Regarding static analysis overhead \cite{kim2024evaluating}, Figure \ref{fig:statitime} demonstrates that \toolname requires preprocessing time comparable to AFLGo and AFLGo-M. Although occasionally slower than WindRanger and SelectFuzz-VT, this one-time offline cost is entirely acceptable given the substantial runtime performance improvements.

\begin{table*}[]
\centering
\caption{Time-to-Expose Results of Ablation}
\vspace{-0.3cm}
\resizebox{0.82\textwidth}{!}{
\renewcommand{\arraystretch}{1.1} 
\begin{tabular}{@{}ll|rr|rrr|rrr|rrr|rrr|rrr@{}}
\toprule
\multicolumn{1}{c|}{\multirow{2}{*}{\textbf{CVE}}} & \multirow{2}{*}{\textbf{No.}} & \multicolumn{2}{c|}{\textbf{\toolname}} & \multicolumn{3}{c|}{\textbf{NO-KF}}& \multicolumn{3}{c|}{\textbf{NO-APM}} & \multicolumn{3}{c|}{\textbf{NO-MOS}} & \multicolumn{3}{c|}{\textbf{NO-TTC}} & \multicolumn{3}{c}{\textbf{AFLGo-VT}} \\ \cmidrule(l){3-19} 
\multicolumn{1}{c|}{}                              &                               & Runs                    & $\mu TTE$                      & Runs   & $\mu TTE$   & $R_{tte}$     & Runs   & $\mu TTE$   & $R_{tte}$   & Runs  & $\mu TTE$      & $R_{tte}$ & Runs  & $\mu TTE$      & $R_{tte}$ & Runs  & $\mu TTE$       & $R_{tte}$  \\ \midrule
\multicolumn{1}{l|}{CVE-2020-36130}                & 1                             & 10                    & {\ul 35.78}                    & 10                       & 36.27                         & 1.01                           & 10    & 68.88           & 1.92      & 10      & 86.89        & 2.43        & 10 & 77.09 & 2.15      & 10     & 206.37          & 5.77       \\
\multicolumn{1}{l|}{CVE-2023-2908}                 & 3                             & 10                    & 1885.48                        & 10                       & {\ul 1842.89}                 & 0.98                           & 10    & 5150.36         & 2.73      & 10      & 4714.61      & 2.50        & 10                      & 6161.04                    & 3.27      & 10     & 7667.62         & 4.07       \\
\multicolumn{1}{l|}{CVE-2023-2908}                 & 4                             & 10                    & 3474.33                        & 10                       & {\ul 2725.49}                 & 0.78                           & 10    & 5546.17         & 1.60      & 10      & 9506.05      & 2.74        & 10                      & 3681.19                    & 1.06      & 10     & 20496.89        & 5.90       \\
\multicolumn{1}{l|}{CVE-2023-2908}                 & 5                             & 10                    & {\ul 1671.75}                  & 10                       & 2202.02                       & 1.32                           & 10    & 18204.76        & 10.89     & 9       & 7377.25      & 4.41        & 9                       & 14173.43                   & 8.48      & 7      & 35280.16        & 21.10      \\
\multicolumn{1}{l|}{CVE-2021-3537}                 & 12                            & 10                    & 10223.66                       & 7                        & {\ul 6376.88}                 & 0.62                           & 4     & 39342.07        & 3.85      & 9       & 29904.87     & 2.93        & 10                      & 32706.08                   & 3.20      & 6      & 49965.42        & 4.89       \\
\multicolumn{1}{l|}{CVE-2021-3517}                 & 12                            & 10                    & {\ul 65.13}                    & 10                       & 73.2                          & 1.12                           & 10    & 126.79          & 1.95      & 10      & 344.99       & 5.30        & 10                      & 256.04                     & 3.93      & 10     & 298.71          & 4.59       \\
\multicolumn{1}{l|}{CVE-2021-3518}                 & 13                            & 10                    & 267.21                         & 10                       & {\ul 239.72}                  & 0.90                           & 9     & 18332.58        & 68.61     & 10      & 11575.65     & 43.32       & 9                       & 2511.95                    & 9.40      & 10     & 15638.62        & 58.52      \\
\multicolumn{1}{l|}{CVE-2020-24977}                & 14                            & 2                     & {\ul 17100.59}                 & 2                        & 48665.94                      & 2.85                           & TO    & TO              & TO        & 1       & 64096.18     & 3.75        & TO                      & TO                         & TO        & 0      & TO              & TO         \\
\multicolumn{1}{l|}{CVE-2021-37616}                & 16                            & 10                    & 631.19                         & 10                       & {\ul 627.92}                  & 0.99                           & 10    & 746.96          & 1.18      & 10      & 810.51       & 1.28        & 10                      & 875.70                     & 1.39      & 10     & 2558.37         & 4.05       \\
\multicolumn{1}{l|}{CVE-2021-34335}                & 16                            & 10                    & 1464.68                        & 10                       & {\ul 563.03}                  & 0.38                           & 9     & 4824.21         & 3.29      & 10      & 1279.54      & 0.87        & 10                      & 5102.24                    & 3.48      & 9      & 9607.61         & 6.56       \\
\multicolumn{1}{l|}{CVE-2021-29457}                & 17                            & 10                    & {\ul 406.78}                   & 10                       & 873.61                        & 2.15                           & 10    & 615.57          & 1.51      & 10      & 1073.91      & 2.64        & 10                      & 665.63                     & 1.64      & 10     & 2387.48         & 5.87       \\
\multicolumn{1}{l|}{CVE-2021-29458}                & 17                            & 10                    & {\ul 1327.31}                  & 10                       & 2988.08                       & 2.25                           & 10    & 2424.48         & 1.83      & 10      & 3013.39      & 2.27        & 10                      & 1819.96                    & 1.37      & 10     & 7234.35         & 5.45       \\
\multicolumn{1}{l|}{CVE-2021-29470}                & 17                            & 10                    & {\ul 539.50}                   & 10                       & 848.26                        & 1.57                           & 10    & 874.64          & 1.62      & 10      & 1454.26      & 2.70        & 10                      & 667.44                     & 1.24      & 10     & 4371.51         & 8.10       \\
\multicolumn{1}{l|}{CVE-2021-37618}                & 18                            & 10                    & {\ul 24.25}                    & 10                       & 31.38                         & 1.29                           & 10    & 71.54           & 2.95      & 10      & 61.87        & 2.55        & 10                      & 46.84                      & 1.93      & 10     & 174.74          & 7.21       \\
\multicolumn{1}{l|}{CVE-2021-37619}                & 19                            & 10                    & {\ul 707.45}                   & 10                       & 820.78                        & 1.16                           & 10    & 1241.53         & 1.75      & 10      & 1388.63      & 1.96        & 10                      & 1429.55                    & 2.02      & 10     & 4433.38         & 6.27       \\
\multicolumn{1}{l|}{CVE-2021-29457}                & 19                            & 10                    & {\ul 24.17}                    & 10                       & 29.1                          & 1.20                           & 10    & 42.41           & 1.75      & 10      & 53.09        & 2.20        & 10                      & 64.43                      & 2.67      & 10     & 132.66          & 5.49       \\
\multicolumn{1}{l|}{CVE-2020-18831}                & 20                            & 10                    & {\ul 56.40}                    & 10                       & 165.93                        & 2.94                           & 10    & 830.12          & 14.72     & 10      & 75.84        & 1.34        & 10                      & 374.56                     & 6.64      & 10     & 2134.72         & 37.85      \\
\multicolumn{1}{l|}{CVE-2021-3478}                 & 21                            & 10                    & {\ul 1048.43}                  & 10                       & 1191.27                       & 1.14                           & 10    & 1202.83         & 1.15      & 10      & 1162.80      & 1.11        & 10                      & 1647.97                    & 1.57      & 10     & 3071.03         & 2.93       \\
\multicolumn{1}{l|}{CVE-2021-20302}                & 21                            & 8                     & 33187.90                       & 7                        & 23826.12                      & 0.72                           & 3     & 57581.06        & 1.74      & 8       & 36858.80     & 1.11        & 2                       & 50257.11                   & 1.51      & 3      & {\ul 8283.01}   & 0.25       \\
\multicolumn{1}{l|}{CVE-2020-11764}                & 21                            & 6                     & 31226.66                       & 4                        & {\ul 31045.86}                & 0.99                           & 3     & 46595.99        & 1.49      & 3       & 50449.67     & 1.62        & 2                       & 56256.76                   & 1.80      & 1      & 37558.00        & 1.20       \\
\multicolumn{1}{l|}{CVE-2020-11765}                & 21                            & 7                     & 13227.39                       & 5                        & 11976.83                      & 0.91                           & 1     & {\ul 9030.95}   & 0.68      & 9       & 19971.73     & 1.51        & 4                       & 16998.51                   & 1.29      & 2      & 35330.87        & 2.67       \\
\multicolumn{1}{l|}{CVE-2020-16587}                & 21                            & 9                     & {\ul 13732.24}                 & 9                        & 31282.23                      & 2.28                           & 6     & 42419.61        & 3.09      & 8       & 28339.87     & 2.06        & 6                       & 47468.85                   & 3.46      & 2      & 18304.84        & 1.33       \\
\multicolumn{1}{l|}{CVE-2021-3474}                 & 21                            & 10                    & {\ul 1006.05}                  & 10                       & 1142.39                       & 1.14                           & 10    & 1324.19         & 1.32      & 10      & 1198.93      & 1.19        & 10                      & 1744.42                    & 1.73      & 10     & 2857.82         & 2.84       \\
\multicolumn{1}{l|}{CVE-2021-23215}                & 22                            & 3                     & 23641.52                       & 1                        & 31894.97                      & 1.35                           & 1     & {\ul 20793.05}  & 0.88      & 1       & 55582.50     & 2.35        & 2                       & 47540.21                   & 2.01      & 0      & TO              & TO         \\
\multicolumn{1}{l|}{CVE-2021-23169}                & 22                            & 10                    & {\ul 1840.21}                  & 10                       & 7229.39                       & 3.93                           & 10    & 4843.77         & 2.63      & 10      & 8533.81      & 4.64        & 10                      & 14983.46                   & 8.14      & 9      & 26141.88        & 14.21      \\
\multicolumn{1}{l|}{CVE-2021-3478}                 & 22                            & 10                    & 114.08                         & 10                       & {\ul 106.54}                  & 0.93                           & 10    & 123.88          & 1.09      & 10      & 235.07       & 2.06        & 10                      & 273.92                     & 2.40      & 10     & 331.37          & 2.90       \\
\multicolumn{1}{l|}{CVE-2021-20302}                & 22                            & 10                    & {\ul 17053.73}                 & 9                        & 31684.65                      & 1.86                           & 6     & 23653.53        & 1.39      & 8       & 34403.36     & 2.02        & 6                       & 26274.57                   & 1.54      & 5      & 49781.32        & 2.92       \\
\multicolumn{1}{l|}{CVE-2020-11765}                & 22                            & 8                     & 4165.56                        & 7                        & {\ul 3161.39}                 & 0.76                           & 7     & 13462.28        & 3.23      & 9       & 7415.29      & 1.78        & 9                       & 24752.50                   & 5.94      & 8      & 24055.42        & 5.77       \\
\multicolumn{1}{l|}{CVE-2020-16587}                & 22                            & 9                     & {\ul 25941.72}                 & 9                        & 26130.13                      & 1.01                           & 8     & 26333.21        & 1.02      & 8       & 34560.62     & 1.33        & 7                       & 26822.63                   & 1.03      & 6      & 43080.77        & 1.66       \\
\multicolumn{1}{c|}{CVE-2021-3474}                 & 22                            & 10                    & 127.51                         & 10                       & {\ul 98.26}                   & 0.77                           & 10    & 198.53          & 1.56      & 10      & 244.24       & 1.92        & 10                      & 238.77                     & 1.87      & 10     & 407.00          & 3.19       \\
\multicolumn{1}{c|}{CVE-2020-11760}                & 22                            & 1                     & {\ul 1806.09}                  & \multicolumn{1}{r}{0}    & \multicolumn{1}{r}{TO}        & \multicolumn{1}{r|}{TO}        & 0     & TO              & -         & 0       & TO           & -           & 0                       & TO                         & -         & 0      & TO              & -          \\ \hline
\multicolumn{2}{c|}{Avg.}                                                          & \multicolumn{2}{l|}{}                                  & \multicolumn{3}{c|}{1.21}                                                                 & \multicolumn{3}{c|}{2.25}           & \multicolumn{3}{c|}{2.30}            & \multicolumn{3}{c|}{2.43}                                        & \multicolumn{3}{c}{4.74}              \\ \hline
\end{tabular}
}
\vspace{-0.3cm}
\label{table:rq3}
\end{table*}

\greyboxb{Summary for RQ2:} {
Compared to baselines, {\toolname} increases the speed of vulnerability exposure by 5.85x, 4.74x, 6.25x, 7.08x, 5.88x, 4.80x, and 5.52x.
In ten runs, {\toolname} triggers 45 more vulnerabilities than the best-performing baseline, and triggers three vulnerabilities that baselines fail to trigger.
}

\subsection{RQ3: Ablation Study}
\label{subsec:rq3}


To answer RQ3, we conduct an ablation study on cases exposing vulnerabilities within 24 hours.
We evaluate four variants: \textbf{NO-TTC} (removes Target Tuple Construction, using Vulnerability Targets similar to AFLGo-VT), \textbf{NO-APM} (removes Abstract Path Mapping, using AFLGo's seed distance), and \textbf{NO-MOS} (removes Mutation Operator Schedule in Section~\ref{ams}).
Moreover, to assess the effectiveness of {\toolname} in scenarios without vulnerability descriptions, we disabled the key function component (i.e., \textbf{NO-KF}).


The results are shown in Table~\ref{table:rq3}. 
We observe that removing any module increases \(\rm \mu TTE\), yet {\toolname} continues to outperform AFLGo-VT, even when a module is disabled.
Specifically, removing the Target Tuple Construction module increases \(\rm \mu TTE\) to 2.43x the original, removing the Abstract Path Mapping module increases \(\rm \mu TTE\) to 2.25x the original, and removing the Mutation Operator Schedule module increases \(\rm \mu TTE\) to 2.30x the original.
When key functional modules are disabled, the \(\rm \mu TTE\) increases to 1.21x.
However, it still significantly outperforms the best-performance baseline, indicating that, even in the absence of specific vulnerability descriptions, {\toolname} is able to maintain robust performance.
Moreover, key functions do not necessarily bring positive improvements.
This is because, for the client program, the triggering of some vulnerabilities may not necessarily require the key function, which confirms that all reachable paths need to be tested as much as possible.
\greyboxb{Summary for RQ3:} {
Removing three modules of {\toolname} individually increases \(\rm \mu TTE\) by 2.43x, 2.25x, and 2.30x, respectively.
This shows that all three modules contribute uniquely to {\toolname}.
Additionally, even in the absence of vulnerability description, {\toolname} still maintains good performance, with \(\rm \mu TTE\) increasing by only 1.21x.
}


\section{Discussion}
\label{sec:discussion}
Here, we discuss the scalability and limitations of {\toolname}.
\subsection{Scalability}  
\noindent \textbf{Language.} Our implementation primarily focuses on C/C++ projects, given their pervasive use in critical software components and open-source supply chains, along with the security risk they bring. 
In theory, the principles of {\toolname} (e.g., the target tuple and the risk-based fuzzing) can also be applied to other language ecosystems that support ICFG analysis (e.g., Java, Rust).
We will extend {\toolname} to more programming languages in future research, further enhancing its impact on software supply chain security.

\noindent \textbf{Deeply Nested Library.} 
To handle deeply nested library calls, {\toolname} can be extended by expanding the 2-tuple target $\langle \texttt{CT}, \texttt{VT} \rangle$ into an n-tuple $\langle \texttt{CT}, \texttt{VT}_1, \texttt{VT}_2, \dots, \texttt{VT}_n \rangle$ to capture intermediate interactions.
Since {\toolname} calculates risk independently for each program, expanding to an n-tuple simply requires calculating and aggregating the risks across all layers.
We leave an implementation and large-scale evaluation of this extension to future work.


\begin{figure}[t]
\centerline{\includegraphics[width=0.85\linewidth]{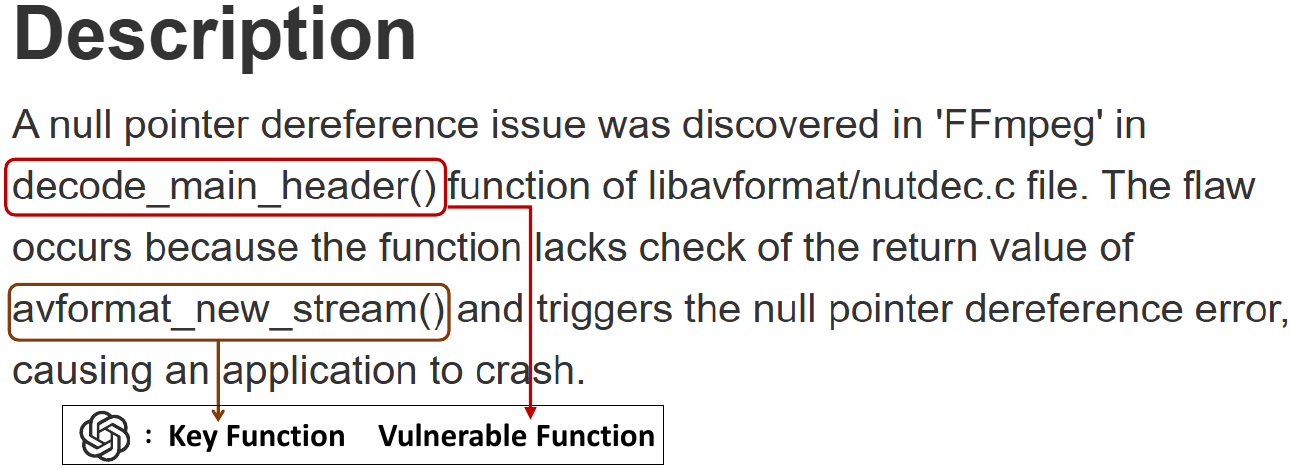}}
\vspace{-0.2cm}
\caption{Extracting information from the vulnerability description of CVE-2022-3341.}
\vspace{-0.3cm}
\label{d1}
\end{figure}
\noindent \textbf{Information Extraction.} 
{\toolname} uses vulnerable and key functions to guide the fuzzer.
While currently requiring manual extraction, our preliminary tests demonstrate that LLMs can infer these functions (e.g., \verb|decode_main_header| and \verb|avformat_new_stream| in Figure~\ref{d1}) from descriptions. Consequently, we plan to integrate an LLM-based extraction in the future.



\subsection{Threats to Validity}  
The first threat is the initial inputs, which significantly impact the effectiveness of fuzzing.
While extensive research explores seed selection, it is not the focus of this paper.
To ensure fairness in the experiments, we used the same seeds to minimize the impact of seed selection on the fuzzing results.
Note that the initial seed does not contain any input that can trigger the vulnerability.

The second threat is the dataset.
We used PoCs to evaluate the exploitability of library vulnerabilities from real client programs, and finally formed a dataset containing 61 such cases.
This approach minimizes the errors introduced by human evaluation and establishes a reliable ground truth for our evaluation.
However, since the vulnerabilities in the dataset can be triggered by existing PoCs, it introduces concerns that the dataset is too simple.
To fit the actual scenario and avoid such concerns, we conduct strict checks when collecting the initial seeds to ensure that no input that can trigger the vulnerability is included in the initial seeds.
In the future, we will continue to collect more client programs to expand our dataset.

The third threat is the reliance on source code and vulnerability reports. 
Like most DGF approaches, {\toolname} requires source code for analysis.
As mentioned in Section~\ref{sec:intro}, we perform exploitability analysis on open-source library vulnerabilities.
Therefore, in our scenario, the source code is available.
Additionally, {\toolname} relies on information in vulnerability reports for auxiliary guidance, which raises concerns about its applicability in situations with limited information.
Compared to existing approaches, which rely on PoCs to run, {\toolname} takes the information in vulnerability reports as an option to further optimize its performance.
Furthermore, ablation study results show that even in the absence of vulnerability description (unable to identify the key function), {\toolname} still maintains significant effectiveness over existing DGF approaches, and improvement over the best-performing baseline by 4.06x.


\section{Related Work}
\textbf{Directed Greybox Fuzzing.} Directed Greybox fuzzing (DGF) enhances the detection of vulnerabilities in target programs by guiding the mutation of seeds.
The distance metric is an important factor in DGF.
AFLGo~\cite{aflgo} is the first line of research about DGF, proposing the concept of distance metrics.
Hawkeye~\cite{chen2018hawkeye} and WindRanger~\cite{du2022windranger} further optimized the distance calculation method.
SDFUZZ~\cite{li2024sdfuzz} and PDGF~\cite{zhang2023predecessor} introduce novel distance metrics (e.g., state feedback or regional maturity).
Titan~\cite{huang2023titan} explores the correlation between multiple vulnerability targets.
Beacon~\cite{huang2022beacon} and SieveFuzz~\cite{srivastava2022one} increase the efficiency of DGF by terminating unreachable paths.
Some recent work (e.g., HGFuzzer~\cite{xu2025directed} ) has integrated LLM to improve the efficiency of DGF.
{\toolname} proposes a new distance calculation to avoid the preference for seeds on shorter paths.
Works such as WindRanger~\cite{du2022windranger} and SelectFuzzz~\cite{luo2023selectfuzz} also try to mitigate shorter path preference.
However, when there are multiple targets, they tend to select seeds closest to a target site rather than those with better overall performance (i.e., close to multiple targets).
In contrast, {\toolname} maps execution paths on different reachable paths to a unified reachable path and still uses the average basic block distances within the execution path.
The average distance can effectively represent the distance of the execution path to multiple targets~\cite{aflgo}.
Experimental results demonstrate our advantages over these works.
Additionally, CAFL~\cite{lee2021constraint} defines a constraint as the combination of a target site and the data conditions.
LOLLY~\cite{liang2019sequence} generates inputs to reach statements and trigger program bugs sequentially.
These studies focus on detecting specific types of vulnerabilities and require inputs to satisfy multiple goal sets in sequence.
Unlike these works, which treat all predecessor nodes as necessary conditions, {\toolname} uses the target tuple to assist in seed distance calculation.
In this design, the \verb|CT| serves as a heuristic guide to the \verb|VT| rather than a strict prerequisite.
Some studies also apply DGF to new scenarios.
FuzzSlice~\cite{murali2024fuzzslice} combines static analysis and DFG to reduce false positives.
1dVul~\cite{peng20191dvul} uses DGF to discover 1-Day vulnerabilities through binary patches.
GrayC~\cite{even2023grayc} proposes a novel greybox fuzzer for C compilers and analyzers.

\noindent \textbf{Greybox Fuzzing.} Greybox Fuzzing (GF) is a widely used testing technology.
SEAMFUZZ~\cite{lee2023learning} clusters seeds based on syntactic and semantic similarities and learns mutation strategies for each cluster using a customized Thompson sampling algorithm.
MOPT~\cite{lyu2019mopt} dynamically evaluates the efficiency of mutation operators for each program and adjusts their selection probabilities toward an optimal distribution.
Deepfuzzer~\cite{liang2019deepfuzzer} leverages lightweight symbolic execution for generating qualified seeds and balances mutation frequency using a statistical seed selection algorithm.
LTL-Fuzzer~\cite{meng2022linear} proposes a new algorithm to find violations of Linear-time Temporal Logic properties based on GF.
Mu2~\cite{laybourn2022guiding} combines incorporating mutation analysis and GF to produce a seed corpus with a high mutation score.
Ba et al.~\cite {ba2022stateful} constructs a state space ``map'' by identifying state variables and tracking their assigned value sequences.
CovRL-Fuzz~\cite{eom2024fuzzing} combines coverage feedback with an LLM-based mutator via reinforcement learning to provide coverage guidance for LLM-based fuzzing.
These studies often focus on improving overall coverage or discovering a specific type of vulnerability, which is different from detecting the library vulnerabilities' exploitability from client programs.

\section{Conclusion and Future Work}
\label{sec:conclusion}

In this paper, we propose {\toolname}, which extends DGF with the target tuple to detect the exploitability of library vulnerabilities from client programs without PoCs. 
{\toolname} proposes a novel Abstract Path Mapping mechanism to fairly compare seeds in different reachable paths and avoid neglecting seeds that are executed along longer reachable paths.
{\toolname} also generates a customized mutation strategy for each seed according to its risk.
We collect a new dataset to evaluate the {\toolname}.
The results show that, compared to baselines, {\toolname} discovers 37\%, 40\%, 45\%, 69\%, 195\%, 155\%, and 51\% more target-reachable paths and exposes vulnerabilities 5.85x, 4.74x, 6.25x, 7.08x, 5.88x, 4.80x, and 5.52x faster.
We trigger the largest number of vulnerabilities in the ten repeated experiments, and three vulnerabilities are triggered only by {\toolname}.
Additionally, ablation studies confirm that even without key functions, {\toolname} still demonstrates significant improvements compared to the best-performing baseline.
In the future, we plan to extend {\toolname} to deeply nested library vulnerabilities by using the n-tuple and to use LLMs to drive further automated testing.
\newpage

\bibliographystyle{ACM-Reference-Format}
\bibliography{references}

\end{document}